\documentclass[journal,twoside,twocolumn,a4paper]{IEEEtran}
\usepackage{amsmath,amssymb,amscd,latexsym,dsfont}
\usepackage{multirow}
\usepackage{float}
\usepackage{comment}
\usepackage{color}
\usepackage{graphicx}
\usepackage{comment}
\usepackage{subfigure}
\usepackage{enumerate}
\usepackage{stfloats}
\usepackage{psfrag}
\usepackage{setspace}
\usepackage{cite}
\usepackage{balance}
\usepackage{algorithmic}
\usepackage[]{algorithm}
\IEEEoverridecommandlockouts
\usepackage[acronym,nonumberlist]{glossaries} 

\makeatletter
\appto\newacronymhook{%
  \newbool{glo@\the\glslabeltok @usedonlyonce}
}

\patchcmd{\@gls@}{%
  \glsunset{#2}%
}{
  \ifglsused{#2}{%
    \write\@auxout{\global\setbool{glo@#2@usedonlyonce}{false}}%
  }{%
    \write\@auxout{\global\setbool{glo@#2@usedonlyonce}{true}}%
  }%
  \glsunset{#2}%
}{}{}

\patchcmd{\@gls@}{%
  \glsentryfirst{#2}%
}{
  \ifbool{glo@#2@usedonlyonce}{\glsentrylong{#2}}{\glsentryfirst{#2}}%
}{}{}
\makeatother

\newacronym[\glsshortpluralkey=AGCs,\glslongpluralkey=anti-Gray codes]{agc}{AGC}{anti-Gray code}
\newacronym[]{ao}{AO}{\emph{asymptotically optimal}}
\newacronym[]{awgn}{AWGN}{additive white Gaussian noise}
\newacronym[]{bep}{BEP}{bit-error probability}
\newacronym[]{bicm}{BICM}{bit-interleaved coded modulation}
\newacronym[]{bicm-gmi}{BICM-GMI}{BICM generalized mutual information}
\newacronym[]{brgc}{BRGC}{binary reflected Gray code}
\newacronym[\glsshortpluralkey=CDFs,\glslongpluralkey=cumulative density functions]{cdf}{CDF}{cumulative density function}
\newacronym[\glsshortpluralkey=LLRs,\glslongpluralkey=logarithmic likelihood ratios]{llr}{LLR}{logarithmic likelihood ratio}
\newacronym[\glsshortpluralkey=EDs,\glslongpluralkey=Euclidean distances]{ed}{ED}{Euclidean distance}
\newacronym[\glsshortpluralkey=GCs,\glslongpluralkey=Gray codes]{gc}{GC}{Gray code}
\newacronym[]{lhs}{l.h.s.}{left-hand side}
\newacronym[]{map}{MAP}{maximum a posteriori probability}
\newacronym[]{me}{ME}{mean estimator}
\newacronym[\glsshortpluralkey=MEDs,\glslongpluralkey=minimum Euclidean distances]{med}{MED}{minimum Euclidean distance}
\newacronym[\glsshortpluralkey=MIs,\glslongpluralkey=mutual informations]{mi}{MI}{mutual information}
\newacronym[]{mmse}{MMSE}{minimum mean-square error}
\newacronym[]{nbc}{NBC}{natural binary code}
\newacronym[\glsshortpluralkey=PDFs,\glslongpluralkey=probability density functions]{pdf}{PDF}{probability density function}
\newacronym[\glsshortpluralkey=PMFs,\glslongpluralkey=probability mass functions]{pmf}{PMF}{probability mass function}
\newacronym[]{rhs}{r.h.s.}{right-hand side}
\newacronym[]{sep}{SEP}{symbol-error probability}
\newacronym[\glsshortpluralkey=SNRs,\glslongpluralkey=signal-to-noise ratios]{snr}{SNR}{signal-to-noise ratio}
\newacronym[]{qam}{QAM}{quadrature amplitude modulation}

\newcommand{\tdelta}{\epsilon}
\newcommand{\MaxED}{\hat{d}}
\newcommand{\MED}{d}

\newcommand{\PMFU}{\PMF^{\tr{u}}}
\newcommand{\PMFEU}{\PMF^{\tr{eu}}}
\newcommand{\A}{A}
\newcommand{\Ax}{\A_{\mcX}}

\newcommand{\Axkb}{\A_{\mcXkb}}
\newcommand{\Axkz}{\A_{\mcXkz}}
\newcommand{\Axko}{\A_{\mcXko}}

\newcommand{\Axkbkb}[1]{\A_{\mc{X}_{#1}}}
\newcommand{\Ae}{\A_{\mcE}}
\newcommand{\B}{B}

\newcommand{\Bxpkb}{\B_{\PMFkb}}
\newcommand{\Bxp}{\B_{\PMF}}
\newcommand{\Bxu}{\B_{\PMFU}}
\newcommand{\C}{C}
\newcommand{\Cxl}{\C_{\mcX,\labeling}}
\newcommand{\Cel}{\C_{\mcE,\labeling}}
\newcommand{\D}{D}
\newcommand{\Dxl}{\D_{\PMF,\labeling}}
\newcommand{\Dxlu}{\D_{\PMFU,\labeling}}

\newcommand{\Rxi}{R_{\PMF}^{(i)}} 

\newcommand{\mcDxi}{\mcD_{\mcX}^{(i)}}

\newcommand{\mcIX}{\mc{I}_{\mcX}}
\newcommand{\setCx}{\mcC_{\mcX}}
\newcommand{\setCe}{\mcC_{\mcE}}
\newcommand{\ENX}{\EN_{\PMF}}
\newcommand{\ENXxpkb}{\EN_{\PMFkb}}
\newcommand{\MIxp}{\MI_{\PMF}(\rho)}
\newcommand{\MIxpkb}{\MI_{\PMFkb}(\rho)}

\newcommand{\MIxu}{\MI_{\PMFU}(\rho)}
\newcommand{\MIeu}{\MI_{\PMFEU}(\rho)}
\newcommand{\MMSExp}{\MMSE_{\PMF}(\rho)}
\newcommand{\MMSExu}{\MMSE_{\PMFU}(\rho)}
\newcommand{\MMSEeu}{\MMSE_{\PMFEU}(\rho)}
\newcommand{\MMSEkpkb}{\MMSE_{\PMFkb}(\rho)}

\newcommand{\textBI}{\textnormal{bicm}}
\newcommand{\textMAP}{\textnormal{map}}
\newcommand{\textME}{\textnormal{me}}
\newcommand{\textMI}{\textnormal{mi}}
\newcommand{\textMMSE}{\textnormal{mmse}}

\newcommand{\MIBIxp}{\MI_{\PMF,\labeling}^{\textBI}(\rho)}
\newcommand{\MIBIxu}{\MI_{\PMFU,\labeling}^{\textBI}(\rho)}
\newcommand{\MMSEBIxp}{\MMSE_{\PMF,\labeling}^{\textBI}(\rho)}
\newcommand{\MMSEBIxu}{\MMSE_{\PMFU,\labeling}^{\textBI}(\rho)}

\newcommand{\KMIxpl}{\K_{\PMF,\labeling}^{\textMI}(\rho)}

\newcommand{\KMIeul}{\K_{\PMFEU,\labeling}^{\textMI}(\rho)}

\newcommand{\KMMSExpl}{\K_{\PMF,\labeling}^{\textMMSE}(\rho)}

\newcommand{\KMMSEeul}{\K_{\PMFEU,\labeling}^{\textMMSE}(\rho)}

\newcommand{\Rxpl}{\R_{\PMF,\labeling}}
\newcommand{\Rxul}{\R_{\mcX,\labeling}}
\newcommand{\Reul}{\R_{\mcE,\labeling}}

\newcommand{\CelAGC}{C_{\mcE,\labeling_{\tr{AGC}}}}
\newcommand{\CelNBC}{C_{\mcE,\labeling_{\tr{NBC}}}}

\newcommand{\ReulBRGC}{\R_{\mcE,\labeling_{\tr{GC}}}}
\newcommand{\ReulAGC}{\R_{\mcE,\labeling_{\tr{AGC}}}}
\newcommand{\ReulNBC}{\R_{\mcE,\labeling_{\tr{NBC}}}}

\newcommand{\pdf}{f}	
\newcommand{\pmf}{P}	
\newcommand{\PMF}{P_{X}}	
\newcommand{\PMFkb}{P_{X_{k,b}}}

\newcommand{\Pxi}{p_i}	
\newcommand{\Pxim}{p_{i-1}}
\newcommand{\Pxip}{p_{i+1}}
\newcommand{\Pxj}{p_j}	

\newcommand{\sumk}{\sum_{k=1}^{m}}	
\newcommand{\sumi}{\sum_{i\in\mcIX}}
\newcommand{\sumj}{\sum_{j\in\mcIX}}
\newcommand{\sumikb}{\sum_{i\in\mcIXkb}}

\newcommand{\sumb}{\sum_{b\in\set{0,1}}}	
\newcommand{\limxinf}[1]{\lim_{x\rightarrow\infty}{#1}}
\newcommand{\limrinf}[1]{{\lim_{\rho\rightarrow\infty}{#1}}}
\newcommand{\limrrinf}[1]{{\lim_{r\rightarrow\infty}{#1}}}
\newcommand{\tr}[1]{\mathrm{#1}}

\newcommand{\mc}[1]{\mathcal{#1}}

\newcommand{\set}[1]{\{#1\}}
\newcommand{\cd}{\cdot}

\newcommand{\ld}{\ldots}

\newcommand{\ms}[1]{\mathds{#1}}

\newcommand{\Real}{\mathbb{R}}

\newcommand{\ie}{i.e.,~}
\newcommand{\eg}{e.g.,~}

\newcounter{lemma}
\newtheorem{theorem}{Theorem}
\newtheorem{corollary}[theorem]{Corollary}
\newtheorem{lemma}[theorem]{Lemma}

\newtheorem{example}{Example}
\newtheorem{remark}{Remark}

\newcommand{\MMSE}[0]{M}
\newcommand{\MI}[0]{I}
\newcommand{\SEP}[0]{S}
\newcommand{\SEPxp}{\SEP_{\PMF}(\rho)}
\newcommand{\SEPxu}{\SEP_{\PMFU}(\rho)}
\newcommand{\SEPeu}{\SEP_{\PMFEU}(\rho)}

\newcommand{\BEP}[0]{B}
\newcommand{\BEPxp}{\BEP_{\PMF,\labeling}(\rho)}
\newcommand{\BEPxu}{\BEP_{\PMFU,\labeling}(\rho)}

\newcommand{\GF}[0]{G}
\newcommand{\QF}[0]{Q}
\newcommand{\EN}[0]{H}
\newcommand{\K}[0]{K}
\newcommand{\R}[0]{T}

\newcommand{\bp}{\boldsymbol{p}}

\newcommand{\AGC}{\mathbf{W}} 

\newcommand{\bl}{\boldsymbol{l}}
\newcommand{\bQ}{\boldsymbol{Q}}
\newcommand{\bq}{\boldsymbol{q}}

\newcommand{\mcC}{\mc{C}}
\newcommand{\mcD}{\mc{D}}
\newcommand{\mcE}{\mc{E}}

\newcommand{\mcX}{\mc{X}}
\newcommand{\mcY}{\mc{Y}}

\newcommand{\mcIXkb}{\mc{I}_{\mc{X}_{k,b}}}
\newcommand{\mcIXkbn}{\mc{I}_{\mc{X}_{k,\overline{b}}}}

\newcommand{\mcXkb}{\mc{X}_{k,b}}
\newcommand{\mcXkz}{\mc{X}_{k,0}}
\newcommand{\mcXko}{\mc{X}_{k,1}}

\newcommand{\mcXkbn}{\mc{X}_{k,\overline{b}}}
\newcommand{\mcIXkz}{\mc{I}_{\mc{X}_{k,0}}}
\newcommand{\mcIXko}{\mc{I}_{\mc{X}_{k,1}}}

\newcommand{\bx}{\boldsymbol{x}}

\newcommand{\by}{{y}}

\newcommand{\argmax}{\mathop{\mathrm{argmax}}}

\newcommand{\Es}{E_\tr{s}}

\newcommand{\Ex}{\ms{E}}
\renewcommand{\exp}[1]{\mathrm{e}^{#1}}
\newcommand{\figref}[1]{Fig.~\ref{#1}}
\newcommand{\tabref}[1]{Table~\ref{#1}}
\newcommand{\secref}[1]{Sec.~\ref{#1}}
\newcommand{\theoref}[1]{Theorem~\ref{#1}}
\newcommand{\lemmaref}[1]{Lemma~\ref{#1}}
\newcommand{\cororef}[1]{Corollary~\ref{#1}}
\newcommand{\exref}[1]{Example~\ref{#1}}
\newcommand{\appref}[1]{Appendix~\ref{#1}}

\newcommand{\T}{^{\mathsf{T}}}		

\newcommand{\labeling}{\bl} 		

\title{High-SNR Asymptotics of Mutual Information for Discrete Constellations with Applications to BICM}
\author{Alex Alvarado, Fredrik Br\"annstr\"om, Erik Agrell, and Tobias Koch
\thanks{Research supported by the European Community's Seventh's Framework Programme (FP7/2007-2013) under grant agreements No.~271986 and No.~333680, by the Swedish Research Council, Sweden (under grants \#621-2006-4872 and \#621-2011-5950) and by the Ministerio de Econom\'ia y Competitividad of Spain (TEC2009-14504-C02-01, CSD2008-00010, and TEC2012-38800-C03-01). This work was presented in part at the IEEE International Symposium on Information Theory, Istanbul, Turkey, July 2013, and at the IEEE Communication Theory Workshop, Phuket, Thailand, June 2013.}
\thanks{A.~Alvarado is with the Dept.~of Engineering, University of Cambridge, Cambridge CB2 1PZ, United Kingdom (email: alex.alvarado@ieee.org).}
\thanks{E.~Agrell and F.~Br\"annstr\"om are with the Dept.~of Signals and Systems, Chalmers Univ.~of Technology, SE-41296 G\"oteborg, Sweden (email: \set{fredrik.brannstrom,\,agrell}@chalmers.se).}
\thanks{T.~Koch is with the Department of Signal Theory and Communications, Universidad Carlos III de Madrid, 28911 Legan\'es, Spain (email: koch@tsc.uc3m.es).}}

\begin{document}
\maketitle

\begin{abstract}
Asymptotic expressions of the mutual information between any discrete input and the corresponding output of the scalar additive white Gaussian noise channel are presented in the limit as the signal-to-noise ratio (SNR) tends to infinity. Asymptotic expressions of the symbol-error probability (SEP) and the minimum mean-square error (MMSE) achieved by estimating the channel input given the channel output are also developed. It is shown that for any input distribution, the conditional entropy of the channel input given the output, MMSE and SEP have an asymptotic behavior proportional to the Gaussian Q-function. The argument of the Q-function depends only on the minimum Euclidean distance (MED) of the constellation and the SNR, and the proportionality constants are functions of the MED and the probabilities of the pairs of constellation points at MED. The developed expressions are then generalized to study the high-SNR behavior of the generalized mutual information (GMI) for bit-interleaved coded modulation (BICM). By means of these asymptotic expressions, the long-standing conjecture that Gray codes are the binary labelings that maximize the BICM-GMI at high SNR is proven. It is further shown that for any equally spaced constellation whose size is a power of two, there always exists an anti-Gray code giving the lowest BICM-GMI at high SNR.
\end{abstract}



\section{Introduction}\label{Sec:Introduction}

\PARstart{W}{e} consider the discrete-time, real-valued, \gls{awgn} channel
\begin{align}\label{AWGN}
Y =\sqrt{\rho}X+Z
\end{align}
where $X$ is the transmitted symbol; $Z$ is a Gaussian random variable, independent of $X$, with zero mean and unit variance; and $\rho>0$ is an arbitrary scale factor. The capacity of the \gls{awgn} channel \eqref{AWGN} under an average-power constraint $\rho \Ex_X[X^2]\leq\gamma$ is given by \cite{Shannon48}
\begin{equation}\label{Capacity}
C^{\tr{aw}}(\gamma) = \frac{1}{2}\log(1+\gamma)
\end{equation}
where $\gamma$ can be viewed as the maximal allowed \gls{snr}. Although inputs distributed according to the Gaussian distribution attain the capacity, they suffer from several drawbacks which prevent them from being used in practical systems. Among them, especially relevant are the unbounded support and the infinite number of bits needed to represent signal points. In practical systems, discrete distributions are typically preferred.

The \gls{mi} between the channel input $X$ and the channel output $Y$ of \eqref{AWGN}, where the distribution of $X$ is constrained to be a \gls{pmf} over a discrete constellation, represents the maximum rate at which information can be reliably transmitted over \eqref{AWGN} using that particular constellation. While the low-\gls{snr} asymptotics of the \gls{mi} for discrete constellations are well understood (see \cite{Shannon48,Verdu90,Verdu02,Prelov04} and references therein), to the best of our knowledge, only upper and lower bounds are known for the high-\gls{snr} behavior \cite{Lozano06,PerezCruz10,Wu11}. It was observed in \cite[p.~1073]{PerezCruz10} that for discrete constellations, maximizing the \gls{mi} is equivalent to minimizing either the \gls{sep} or the \gls{mmse} in estimating $X$ from $Y$. In \cite[App.~E]{Duyck12}, two constellations with different \glspl{med} are compared, and it is shown that, for sufficiently large \gls{snr}, the constellation with larger \gls{med} gives a higher \gls{mi}. Upper and lower bounds on the \gls{mi} and \gls{mmse} for multiple-antenna systems over fading channels can be found in \cite{Rodrigues11a,Rodrigues12a,Rodrigues12c}. Using the Mellin transform, asymptotic expansions for the \gls{mmse} and \gls{mi} for scalar and vectorial coherent fading channels were recently derived in \cite{Rodrigues12b}.

In this paper, we study the high-\gls{snr} asymptotics of the \gls{mi} for discrete constellations. In particular, we consider arbitrary constellations and input distributions (independent of $\rho$) and find exact asymptotic expressions for the \gls{mi} in the limit as the \gls{snr} tends to infinity. Exact asymptotic expressions for the \gls{mmse} and \gls{sep} are also developed. We prove that for any constellation and input distribution, the conditional entropy of $X$ given $Y$, the \gls{mmse}, and the \gls{sep} have an asymptotic behavior proportional to $\QF\left({\sqrt{\rho}\MED}/{2}\right)$, where $\QF(\cdot)$ is the Gaussian Q-function and $\MED$ is the \gls{med} of the constellation. While this asymptotic behavior has been demonstrated for uniform input distributions (\eg \cite[eqs.~(36)--(37)]{PerezCruz10}, \cite[Sec.~II-C]{PerezCruz10}, \cite[Sec.~III]{Rodrigues11a}, \cite[Sec.~III]{Rodrigues12c}), we show that it holds for any discrete input distribution that does not depend on the \gls{snr}. Furthermore, in contrast to previous works, we provide closed-form expressions for the coefficients in front of the Q-functions, thereby characterizing the asymptotic behavior of the \gls{mi}, \gls{mmse}, and \gls{sep} more accurately.

While these asymptotic results are general, we use them to study \gls{bicm} \cite{Zehavi92,Caire98,Fabregas08_Book}, which can be viewed as a pragmatic approach for coded modulation \cite[Ch.~1]{Fabregas08_Book}. The key element in \gls{bicm} is the use of a (suboptimal) bit-wise detection rule, which was cast in \cite{Martinez09} as a mismatched decoder. \gls{bicm} is used in many of the current wireless communications standards, \eg HSPA, IEEE 802.11a/g/n, and the DVB standards (DVB-T2/S2/C2).

The \gls{bicm-gmi} is an achievable rate for \gls{bicm} \cite{Martinez09} and depends heavily on the binary labeling of the constellation. The optimality of a \gls{gc} in the sense that it maximizes the \gls{bicm-gmi} was conjectured in \cite[Sec.~III-C]{Caire98}; however, it was shown in \cite{Stierstorfer07a} that for low and medium \glspl{snr}, there exist other labelings that give a higher \gls{bicm-gmi} (see also \cite[Ch.~3]{Stierstorfer09_Thesis}). For further results on \gls{bicm} at low \gls{snr} see \cite{Martinez08b,Stierstorfer09a,Agrell10b,Agrell12c}. On the other hand, numerical results presented in \cite[Ch.~3]{Stierstorfer09_Thesis} and \cite{Alvarado11b} suggest that \glspl{gc} are indeed optimal at high \gls{snr} in terms of \gls{bicm-gmi}. However, to the best of our knowledge, the optimality of GCs at high SNR has never been proven.

In this paper, we derive an asymptotic expression for the \gls{bicm-gmi} as a function of the constellation, input distribution, and binary labeling. Using this expression, we then prove the optimality of \glspl{gc} at high \gls{snr}. Using the \gls{mi}-\gls{mmse} relationship, an asymptotic expression for the derivative of the \gls{bicm-gmi} is also developed. The obtained asymptotic expressions for the \gls{bicm-gmi} and its derivative, as well as the one for the \gls{bep}, are all shown to converge to their asymptotes proportionally to $\QF\left({\sqrt{\rho}\MED}/{2}\right)$.

This paper is organized as follows. In \secref{Sec:Preliminaries}, the notation convention and system model are presented. The asymptotics of the \gls{mi} and \gls{mmse} are presented in \secref{Sec:Asymptotics} and \gls{bicm} is studied in \secref{Sec:BICM}. The conclusions are drawn in \secref{Sec:Conclusions}.

\section{Preliminaries}\label{Sec:Preliminaries}

\subsection{Notation Convention}\label{Convention}

Row vectors are denoted by boldface letters $\bx=[x_1,x_2,\ld,x_M]$ and sets are denoted by calligraphic letters $\mc{X}$. An exception is the set of real numbers, which is denoted by $\Real$. 
All the logarithms are natural logarithms and all the \glspl{mi} are therefore given in nats. \Glspl{pdf} and conditional \glspl{pdf} are denoted by $\pdf_{Y}(y)$ and $\pdf_{Y|X}(y|x)$, respectively. Analogously, \glspl{pmf} are denoted by $\pmf_{X}(x)$ and $\pmf_{X|Y}(x|y)$. Expectations over a random variable $X$ are denoted by $\Ex_X[\cd]$.

\subsection{Model}\label{Model}

We consider the discrete-time, real-valued \gls{awgn} channel \eqref{AWGN}. The transmitted symbols $X$ belong to a real, one-dimensional \emph{constellation} $\mcX\triangleq\set{x_1,x_2,\ld,x_{M}}$ where $M=2^m$, and they are, without loss of generality, assumed to be distinct and ordered, \ie $x_1<x_2<\cdots<x_M$. Each symbol is transmitted with probability $\Pxi\triangleq \pmf_X(x_i)$, $0< \Pxi <1$, and the vector of probabilities $\bp\triangleq[p_1,\ld,p_M]$ is called the \emph{input distribution}. We assume that neither the constellation nor the input distribution depends on $\rho$.
We denote the set of indices in $\mcX$ and $\bp$ with $\mcIX\triangleq\set{1,\ld,M}$.

The transmitted average symbol energy is
\begin{equation}\label{Es}
\Es\triangleq \Ex_X[X^2] = \sum_{i\in\mcIX} \Pxi x_i^2.
\end{equation}
It follows that the \gls{snr} is $\gamma=\rho\Es$.

An $M$-ary pulse-amplitude modulation ($M$PAM) constellation having $M$ equally spaced symbols (separated by $2\Delta$) is denoted by $\mcE\triangleq\set{x_i=-(M-2i+1)\Delta: i=1,\ld,M}$. A uniform distribution of $X$ is denoted by $\PMFU$, \ie $\Pxi=1/M$ $\forall i$. A uniform input distribution with $\mcX=\mcE$ is denoted by $\PMFEU$, where in this case $\Delta^2=3\Es/(M^2-1)$.

The Gaussian Q-function is defined as
\begin{align}\label{Qfunc}
\QF(x) \triangleq \frac{1}{\sqrt{2\pi}} \int_{x}^{\infty} \exp{-\frac{1}{2} \xi^2}\,\tr{d}\xi.
\end{align}
The entropy of the random variable $X$ is defined as
\begin{equation}
\ENX \triangleq -\Ex_X\left[\log\left(\pmf_{X}(X)\right)\right]
\end{equation}
the \gls{mi} between $X$ and $Y$, $I(X;Y)$, as
\begin{equation} \label{mi-def}
\MIxp\triangleq \Ex_{X,Y}\left[\log\left({\pdf_{Y|X}(Y|X)}/{\pdf_{Y}(Y)}\right)\right]
\end{equation}
and the \gls{mmse} as
\begin{equation} \label{mmse-def}
\MMSExp \triangleq \Ex_{X,Y}[(X-\hat{X}^{\textME}(Y))^2]
\end{equation}
where $\hat{X}^{\textME}(y)\triangleq \Ex_X[X|Y=y]$ is the conditional (posterior) \gls{me}.
We further define the \gls{sep} as
\begin{align}\label{sep}
\SEPxp \triangleq \Pr\set{\hat{X}^{\textMAP}(Y)\ne X}
\end{align}
where $X$ is the transmitted symbol and
\begin{align}\label{map}
\hat{X}^{\textMAP}(y)\triangleq \argmax_{x\in\mcX}\pmf_{X|Y}(x|y)
\end{align}
is the decision made by a \gls{map} symbol demapper.


The \gls{med} of the constellation is defined as
\begin{align}\label{MED}
\MED\triangleq \min_{i,j\in\mcIX:i\neq j}|x_i-x_j|.
\end{align}
We further define $\Ax$ as twice the number of pairs of constellation points at \gls{med}, \ie
\begin{align}
\label{A}
\Ax 
& \triangleq\sumi\sum_{\substack{j\in\mcIX\\|x_{i}-x_{j}|=\MED}} 1.
\end{align}
By using the fact that for any real-valued constellation there are at least one and at most $M-1$ pairs of constellation points at \gls{med}, we obtain the bound
\begin{align}\label{A.Bounds}
2 \leq \Ax \leq 2(M-1).
\end{align}
The upper bound is achieved by an $M$PAM constellation, for which
\begin{align}\label{A.Uniform}
\Ae=2(M-1).
\end{align}
Finally, for a given $\PMF$, we define the constant
\begin{align}
\label{B}
\Bxp 
& \triangleq\sumi\sum_{\substack{j\in\mcIX\\|x_{i}-x_{j}|=\MED}} \sqrt{\Pxj\Pxi}.
\end{align}
For a uniform input distribution, $\PMF=\PMFU$, and thus, 
\begin{align}
\Bxu = \frac{\Ax}{M}.\label{B.A.Uniform}
\end{align}

\begin{example}\label{NES.4PAM.Ex}
Consider an unequally spaced $4$-ary constellation with $x_1=-4$, $x_2=-2$, $x_3=2$, and $x_4=4$, and the input distribution $\Pxi=i/10$ with $i=1,2,3,4$. The \gls{med} in \eqref{MED} is $\MED=2$, $\Es$ in \eqref{Es} is $\Es=10$, $\Ax$ in \eqref{A} is $\Ax=4$ (two pairs of constellation points at \gls{med}), and $\Bxp$ in \eqref{B} is $\Bxp=2\sqrt{p_1 p_2}+2\sqrt{p_3 p_4} \approx 0.98$. 
\end{example}

\section[]{High-\gls{snr} Asymptotics}\label{Sec:Asymptotics}

There exists a fundamental relationship between the \gls{mi} and the \gls{mmse} for \gls{awgn} channels \cite{Guo05} (see also \cite[Ch.~2]{Guo04_Thesis}):
\begin{align}\label{dMI_MMSE}
\frac{\tr{d}}{\tr{d}\rho}\MIxp=\frac{1}{2}\MMSExp.
\end{align}
Exploiting this \gls{mi}-\gls{mmse} relation, bounds on the \gls{mi} can be used to derive bounds on the \gls{mmse} and \emph{vice versa}. The relationship in \eqref{dMI_MMSE} will be used in the proof of \theoref{MI.Gen.Asym.Theo}.

Upper and lower bounds on the \gls{mi} and \gls{mmse} for discrete constellations at high \gls{snr} can be found, \eg in \cite{Lozano06,PerezCruz10,Wu11,Rodrigues11a,Rodrigues12a,Rodrigues12b,Rodrigues12c}. While these bounds describe the correct asymptotic behavior, they are, in general, not tight in the sense that the ratio between them does not tend to one as $\rho\rightarrow\infty$. In what follows, we present exact asymptotic expressions for the \gls{mi} and \gls{mmse} for any arbitrary $\PMF$. We further present exact asymptotic expressions for the \gls{sep}.

\subsection[]{Asymptotics of the \gls{mi}, \gls{mmse}, and \gls{sep}}\label{Asym.MI.and.MMSE}

For any given input distribution $\PMF$, the \gls{mi} tends to $\ENX$ as $\rho$ tends to infinity. In the following we study how fast the \gls{mi} converges towards its maximum $\ENX$ by analyzing the difference $\ENX-\MIxp$, \ie by analyzing the conditional entropy of $X$ given $Y$. \theoref{MI.Gen.Asym.Theo} is the main result of this paper and characterizes the high-\gls{snr} behavior of the conditional entropy $H(X|Y)=\ENX-\MIxp$.

\begin{theorem}\label{MI.Gen.Asym.Theo}
For any $\PMF$
\begin{align}\label{MI.Gen.Asymptotic}
\limrinf{
\frac{\ENX-\MIxp}{\QF\left({\sqrt{\rho}\MED}/{2}\right)}
}
= \pi \Bxp
\end{align}
where $\Bxp$ is given by \eqref{B}.
\end{theorem}
\begin{IEEEproof}
See \appref{MI.Gen.Asym.Theo.Proof}.
\end{IEEEproof}

Similar to \theoref{MI.Gen.Asym.Theo}, we have the following asymptotic expressions for the \gls{mmse} and the \gls{sep}.
\begin{theorem}\label{MMSE.Gen.Asym.Theo}
For any $\PMF$
\begin{align}\label{MMSE.Gen.Asymptotic}
\limrinf{
\frac{\MMSExp}{\QF\left({\sqrt{\rho}\MED}/{2}\right)}
}
= \frac{\pi\MED^2}{4} \Bxp.
\end{align}
\end{theorem}
\begin{IEEEproof}
See \appref{MMSE.Gen.Asym.Theo.Proof}.
\end{IEEEproof}

\begin{theorem}\label{SEP.Gen.Asym.Theo}
For any $\PMF$
\begin{align}\label{SEP.Gen.Asymptotic}
\limrinf{
\frac{\SEPxp}{\QF\left({\sqrt{\rho}\MED}/{2}\right)}
}
= \Bxp.
\end{align}
\end{theorem}
\begin{IEEEproof}
See \appref{SEP.Gen.Asym.Theo.Proof}.
\end{IEEEproof}

Theorems~\ref{MI.Gen.Asym.Theo}--\ref{SEP.Gen.Asym.Theo} reveal that, at high \gls{snr}, the \gls{mi}, \gls{mmse}, and \gls{sep} behave as
\begin{align}
\label{MI.approx}
\MIxp &\approx \ENX - \pi \Bxp\QF\left({\sqrt{\rho}\MED}/{2}\right)\\
\label{MMSE.approx}
\MMSExp &\approx \frac{\pi\MED^2}{4} \Bxp\QF\left({\sqrt{\rho}\MED}/{2}\right)\\
\label{SEP.approx}
\SEPxp &\approx \Bxp\QF\left({\sqrt{\rho}\MED}/{2}\right).
\end{align}
The results in \eqref{MI.approx}--\eqref{SEP.approx} show that for any input distribution, the conditional entropy, MMSE, and SEP have the same high-\gls{snr} behavior: i.e., they are all proportional to a Gaussian Q-function, where the proportionality constants depend on the input distribution and, in the case of the \gls{mmse}, also on the \gls{med} of the constellation. 

\begin{remark}
While the results presented in this section were derived only for one-dimensional (real-valued) constellations, they directly generalize to multidimensional constellations that are constructed as \emph{ordered direct products} \cite[eq.~(1)]{Agrell10b} of one-dimensional constellations. For example, the results directly generalize to rectangular quadrature amplitude modulation constellations.
\end{remark}

\subsection{Discussion and Examples}

For a uniform input distribution ($\PMF=\PMFU$), Theorems~\ref{MI.Gen.Asym.Theo}--\ref{SEP.Gen.Asym.Theo} particularize to the following result.

\begin{corollary}\label{Uniform.Asym.Coro}
For any $\mcX$ and a uniform input distribution
\begin{align}
\label{MI.Uniform.Asymptotic}
\limrinf{
\frac{\log{M}-\MIxu}{\QF\left({\sqrt{\rho}\MED}/{2}\right)}
}
& = \pi \frac{\Ax}{M}\\
\label{MMSE.Uniform.Asymptotic}
\limrinf{
\frac{\MMSExu}{\QF\left({\sqrt{\rho}\MED}/{2}\right)}
}
& = \frac{\pi\MED^2}{4}\frac{\Ax}{M}\\
\label{SEP.Uniform.Asymptotic}
\limrinf{
\frac{\SEPxu}{\QF\left({\sqrt{\rho}\MED}/{2}\right)}
}
&= \frac{\Ax}{M}
\end{align}
where $\Ax$ is given by \eqref{A}.
\end{corollary}
\begin{IEEEproof}
From Theorems~\ref{MI.Gen.Asym.Theo}--\ref{SEP.Gen.Asym.Theo} and \eqref{B.A.Uniform}.
\end{IEEEproof}

The expression \eqref{SEP.Uniform.Asymptotic} corresponds to the well-known high-\gls{snr} approximation for the \gls{sep} \cite[eq.~(2.3-29)]{Anderson03_Book}. \cororef{Uniform.Asym.Coro} shows that for a uniform input distribution, the \gls{mi}, the \gls{mmse}, and the \gls{sep} for discrete constellations in the high-\gls{snr} regime are functions of the MED of the constellation and the number of pairs of constellation points at MED.

For $M$PAM and a uniform input distribution ($\PMF=\PMFEU$), it follows from \cororef{Uniform.Asym.Coro} and \eqref{A.Uniform} that
\begin{align}\label{MI.Uniform.PAM.Asymptotic}
\MIeu &\approx \log{M}-\frac{2\pi(M-1)}{M} \QF\left({\sqrt{\rho}\MED}/{2}\right) \\
\MMSEeu &\approx \frac{6\pi\Es}{M(M+1)} \QF\left({\sqrt{\rho}\MED}/{2}\right) \label{MMSE.Uniform.PAM.Asymptotic} \\
\SEPeu &\approx \frac{2(M-1)}{M} \QF\left({\sqrt{\rho}\MED}/{2}\right). \label{SEP.Uniform.PAM.Asymptotic}
\end{align}
\tabref{Table.Summary} summarizes the results obtained in Theorems~\ref{MI.Gen.Asym.Theo}--\ref{SEP.Gen.Asym.Theo}, \cororef{Uniform.Asym.Coro}, and \eqref{MI.Uniform.PAM.Asymptotic}--\eqref{SEP.Uniform.PAM.Asymptotic}.

\begin{table}[t]
\renewcommand{\arraystretch}{2.3}
\caption[]{Summary of asymptotics of \gls{mi}, \gls{mmse}, and \gls{sep}.}\label{Table.Summary}
\centering
\small
\begin{tabular}{c|ccc}
\hline

\hline
Input Distribution 				& $\PMF$						& $\PMFU$							& $\PMFEU$ \\
\hline
\hline
$\underset{\rho\rightarrow\infty}{\lim}{
\dfrac{\ENX-\MIxp}{\QF\left({\sqrt{\rho}\MED}/{2}\right)}
}$		& $\pi \Bxp$					& $\pi \dfrac{\Ax}{M}$					& $\dfrac{2\pi(M-1)}{M}$\\
\hline
$\underset{\rho\rightarrow\infty}{\lim}{
\dfrac{\MMSExp}{\QF\left({\sqrt{\rho}\MED}/{2}\right)}
}$		& $\dfrac{\pi\MED^2}{4} \Bxp$		& $\dfrac{\pi\MED^2}{4}\dfrac{\Ax}{M}$		& $\dfrac{6\pi\Es}{M(M+1)}$\\
\hline
$\underset{\rho\rightarrow\infty}{\lim}{
\dfrac{\SEPxp}{\QF\left({\sqrt{\rho}\MED}/{2}\right)}
}$		& $\Bxp$		& $\dfrac{\Ax}{M}$	& $\dfrac{2(M-1)}{M}$\\
\hline

\hline
\end{tabular}
\end{table}

\begin{example}\label{Example.4PAM.MMSE}
In \figref{asymptotics_MI_MPAM_IT_rhodB}, we show $\log{M}-\MIeu$ for $4$PAM and $16$PAM with uniform input distributions\footnote{Calculated numerically using Gauss--Hermite quadratures \cite[Sec.~III]{Alvarado11b} with $300$ quadrature points.} together with the asymptotic expression in \eqref{MI.Uniform.PAM.Asymptotic}. We also show the lower and upper bounds derived in \cite[eq.~(34)--(35)]{PerezCruz10} and \cite[eq.~(17)--(19)]{Rodrigues12c}. Observe that \eqref{MI.Uniform.PAM.Asymptotic} approximates $\MIeu$ accurately for a large range of \gls{snr}.
\begin{figure}[tpb]
\newcommand{\scale}{0.8}
\centering
\psfrag{xlabel}[cc][cB][\scale]{$\rho$~[dB]}%
\psfrag{ylabel}[cc][ct][\scale]{$\log{M}-\MIeu$~[nats/symbol]}%
\psfrag{MInfo}[cl][cl][\scale]{Exact}%
\psfrag{MInfoA1}[cl][cl][\scale]{Asymptotic}%
\psfrag{LBFPC}[cl][cl][\scale]{LB \cite{PerezCruz10}}%
\psfrag{UBFPC}[cl][cl][\scale]{UB \cite{PerezCruz10}}%
\psfrag{LBMRDRod}[cl][cl][\scale]{LB \cite{Rodrigues12c}}%
\psfrag{UBMRDRod}[cl][cl][\scale]{UB \cite{Rodrigues12c}}%
\psfrag{4PAM}[cl][cl][\scale]{{\fcolorbox{white}{white}{{\hspace{-0.1cm}$4$PAM}}}}%
\psfrag{16PAM}[cr][cr][\scale]{{\fcolorbox{white}{white}{{$16$PAM\hspace{-0.1cm}}}}}%
\includegraphics[width=0.97\columnwidth]{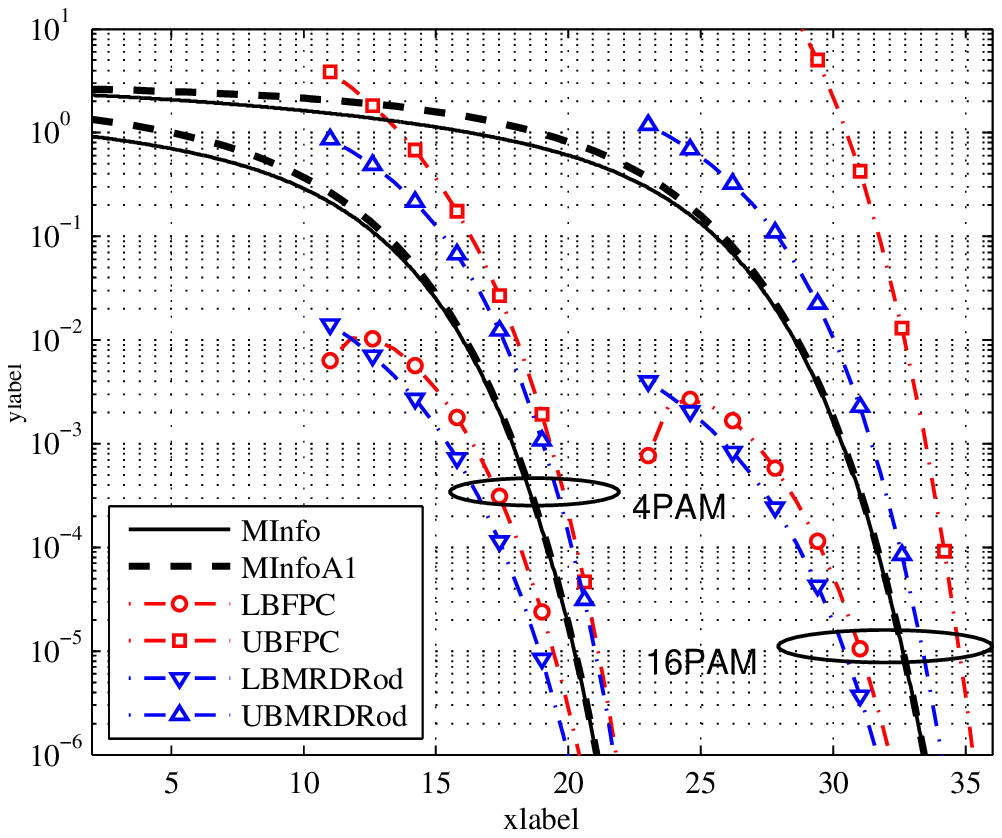}
\caption{$\log{M}-\MIeu$ for $4$PAM and $16$PAM (solid lines) constellations (normalized to $\Es=1$) and the asymptotic expression in \eqref{MI.Uniform.PAM.Asymptotic} (thick dashed lines). The lower and upper bounds \cite[eq.~(34)--(35)]{PerezCruz10} and \cite[eq.~(17)--(19)]{Rodrigues12c} are also shown.}
\label{asymptotics_MI_MPAM_IT_rhodB}
\end{figure}
In \figref{asymptotics_MMSE_MPAM_IT_rhodB}, analogous results for the \gls{mmse} are presented, where the bounds derived in \cite[eq.~(30)--(31)]{PerezCruz10} and \cite[eq.~(13)--(15)]{Rodrigues12c} are also included. Again, the asymptotic expression \eqref{MMSE.Uniform.PAM.Asymptotic} approximates the \gls{mmse} accurately for a large range of \gls{snr}. For other examples with unequally spaced $4$-ary constellations and nonuniform input distributions see \cite[Example 3]{Alvarado13a}.
\begin{figure}[tpb]
\newcommand{\scale}{0.8}
\centering
\psfrag{xlabel}[cc][cB][\scale]{$\rho$~[dB]}%
\psfrag{ylabel}[cc][ct][\scale]{$\MMSEeu$}%
\psfrag{MMSE}[cl][cl][\scale]{Exact}%
\psfrag{MMSEAsy1}[cl][cl][\scale]{Asymptotic}%
\psfrag{LBFPC}[cl][cl][\scale]{LB \cite{PerezCruz10}}%
\psfrag{UBFPC}[cl][cl][\scale]{UB \cite{PerezCruz10}}%
\psfrag{LBMRDRod}[cl][cl][\scale]{LB \cite{Rodrigues12c}}%
\psfrag{UBMRDRod}[cl][cl][\scale]{UB \cite{Rodrigues12c}}%
\psfrag{4PAM}[cl][cl][\scale]{{\fcolorbox{white}{white}{{\hspace{-0.1cm}$4$PAM}}}}%
\psfrag{16PAM}[cr][cr][\scale]{{\fcolorbox{white}{white}{{$16$PAM\hspace{-0.1cm}}}}}%
\includegraphics[width=0.97\columnwidth]{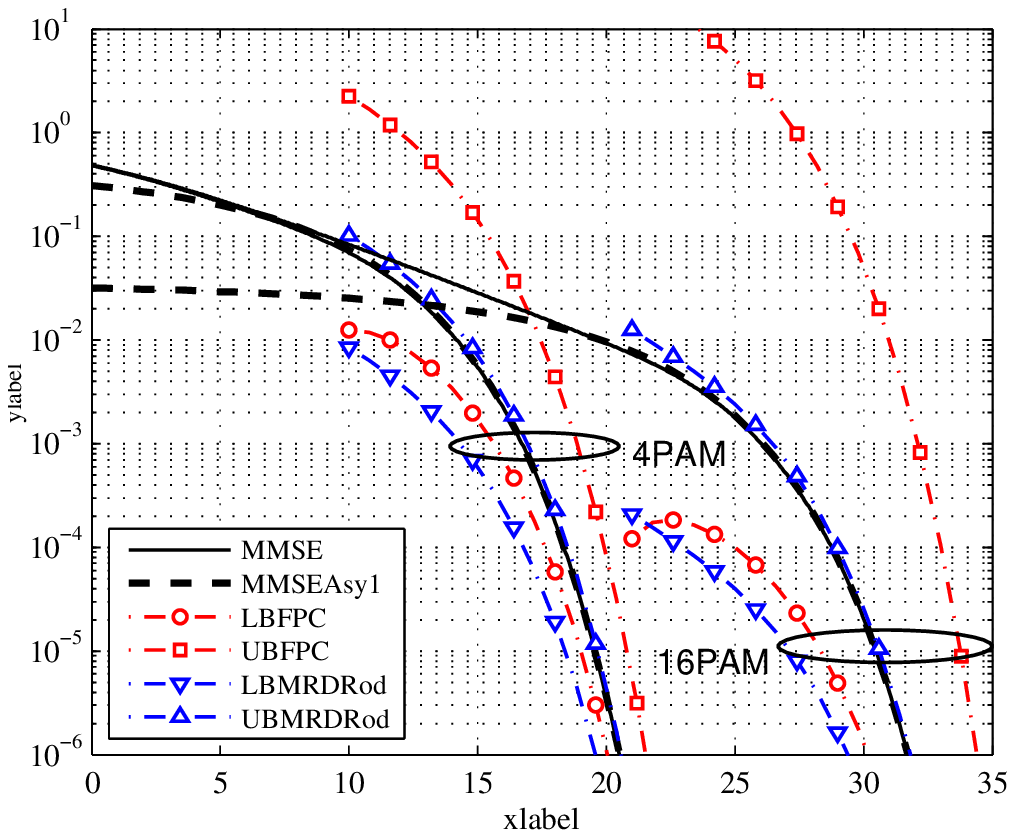}
\caption{$\MMSEeu$ for $4$PAM and $16$PAM (solid lines) constellations (normalized to $\Es=1$) and the asymptotic expression in \eqref{MMSE.Uniform.PAM.Asymptotic} (thick dashed lines). The lower and upper bounds \cite[eq.~(30)--(31)]{PerezCruz10} and \cite[eq.~(13)--(15)]{Rodrigues12c} are also shown.}
\label{asymptotics_MMSE_MPAM_IT_rhodB}
\end{figure}
\end{example}

\begin{remark}
It follows from \cororef{Uniform.Asym.Coro} that the constellation that maximizes the \gls{mi} (or equivalently, the constellation that minimizes the \gls{mmse} and the \gls{sep}) at high \gls{snr} is the constellation that first maximizes the MED and then minimizes $\Ax$. It is easy to see that the one-dimensional constellation that maximizes the \gls{med} for a given $\Es$ is the $M$PAM constellation ($\mcX=\mcE$).
\end{remark}


\begin{figure*}[t]
\newcommand{\scale}{0.95}
\psfrag{Transmitter}[lB][lB][\scale]{BICM Encoder}%
\psfrag{w}[br][Br][\scale]{}%
\psfrag{cpi}[bc][Bc][\scale]{}%
\psfrag{c}[bc][Bc][\scale]{${\bQ}$}%
\psfrag{ENC}[cc][cc][\scale]{ENC}%
\psfrag{PI}[cc][cc][\scale]{$\Pi$}%
\psfrag{MU}[cc][cc][\scale]{$\Phi$}%
\psfrag{sn}[bl][Bl][\scale]{$X$}%
\psfrag{snr}[bc][Bc][\scale]{$\sqrt{\rho}$}%
\psfrag{zn}[bc][Bc][\scale]{${Z}$}%
\psfrag{yn}[br][Br][\scale]{${Y}$}%
\psfrag{MU2}[cc][cc][\scale]{$\Phi^{-1}$}%
\psfrag{PI2}[cc][cc][\scale]{$\Pi^{-1}$}%
\psfrag{DEC}[cc][cc][\scale]{DEC}%
\psfrag{Lkp}[bc][Bc][\scale]{$\boldsymbol{\Lambda}$}%
\psfrag{Lk}[bc][Bc][\scale]{}%
\psfrag{Receiver}[rB][rB][\scale]{BICM Decoder}%
\psfrag{wh}[bl][Bl][\scale]{}%
\begin{center}
	\includegraphics[width=0.75\textwidth]{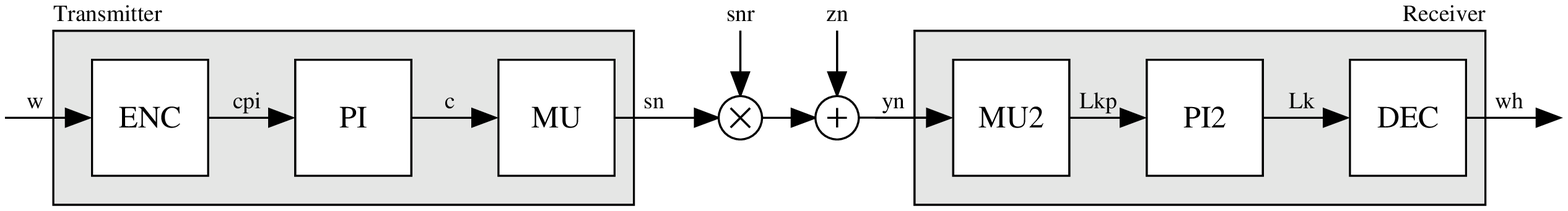}
	\caption{A BICM scheme: The BICM encoder is formed by a serial concatenation of a binary encoder (ENC), a bit-level interleaver ($\Pi$), and a memoryless mapper ($\Phi$). The BICM decoder is based on a demapper ($\Phi^{-1}$) that computes logarithmic likelihood ratios, a deinterleaver ($\Pi^{-1}$), and a channel decoder (DEC).}
	\label{BICM_Model}
\end{center}
\end{figure*}

\section{Application: Binary Labelings for Bit-Interleaved Coded Modulation}\label{Sec:BICM}

\gls{bicm} can be viewed as a pragmatic approach for coded modulation. In \gls{bicm} (see \figref{BICM_Model}), the encoder is realized as a serial concatenation of a binary encoder, a bit-level interleaver, and a memoryless mapper; see \cite{Zehavi92,Caire98,Fabregas08_Book} for more details. A key element in \gls{bicm} is the memoryless mapper $\Phi:\set{0,1}^m\rightarrow\mcX$, which maps the coded bits $\bQ=[Q_{1},\ld,Q_{m}]$ to constellation symbols. At the receiver side, the demapper computes a bit metric, typically in the form of a logarithmic likelihood ratio (LLR)
\begin{align}\label{LLR}
\Lambda_{k} \triangleq \log{\frac{\pdf_{Y|Q_{k}}(y|1)}{\pdf_{Y|Q_{k}}(y|0)}}
\end{align}
for $k=1,\ld,m$. The vector of LLRs $\boldsymbol{\Lambda}=[\Lambda_{1},\ld,\Lambda_{m}]$ is deinterleaved and then decoded.

The \gls{bicm-gmi} is an achievable rate for \gls{bicm}, and thus, is an important quantity for such systems. In this section, we generalize the results in \secref{Sec:Asymptotics} to obtain asymptotic expressions for the \gls{bicm-gmi}. We further study the relationship between the \gls{bicm-gmi} and the \gls{bep} as well as the derivative of the \gls{bicm-gmi} with respect to $\rho$. Finally, we show that at high \gls{snr}, \glspl{gc} maximize \gls{bicm-gmi} for one-dimensional constellations and uniform input distributions.

\subsection{BICM Model}\label{Sec:BICM.Model}

A binary labeling for a constellation is defined by the vector $\labeling=[l_1,l_2,\ld,l_M]$ where $l_i\in\set{0,1,\ld,M-1}$ is the integer representation of the $i$th length-$m$ binary label $\bq_i=[q_{i,1},\ld,q_{i,m}]\in\set{0,1}^{m}$ associated with the symbol $x_i$, with $q_{i,1}$ being the most significant bit. The labeling defines $2m$ \emph{subconstellations} $\mcXkb\subset\mcX$ for $k=1,\ld,m$ and $b\in\set{0,1}$, given by $\mcXkb\triangleq\set{x_i\in\mcX:q_{i,k}=b}$ with $|\mcXkb|=M/2$. We define $\mcIXkb\subset\set{1,\ld,M}$ as the indices of the symbols in $\mcX$ that belong to $\mcXkb$.

\begin{figure}[t]
\newcommand{\scale}{0.8}
\psfrag{x1}[cb][cb][\scale]{$x_1$}\psfrag{x2}[cb][cb][\scale]{$x_2$}
\psfrag{x3}[cb][cb][\scale]{$x_3$}\psfrag{x4}[cb][cb][\scale]{$x_4$}
\psfrag{x5}[cb][cb][\scale]{$x_5$}\psfrag{x6}[cb][cb][\scale]{$x_6$}
\psfrag{x7}[cb][cb][\scale]{$x_7$}\psfrag{x8}[cb][cb][\scale]{$x_8$}
\psfrag{a1}[cb][cb][\scale]{${\boldsymbol{0}}00$}
\psfrag{a2}[cb][cb][\scale]{${\boldsymbol{0}}01$}
\psfrag{a3}[cb][cb][\scale]{${\boldsymbol{0}}10$}
\psfrag{a4}[cb][cb][\scale]{${\boldsymbol{0}}11$}
\psfrag{a5}[cb][cb][\scale]{${\boldsymbol{1}}00$}
\psfrag{a6}[cb][cb][\scale]{${\boldsymbol{1}}01$}
\psfrag{a7}[cb][cb][\scale]{${\boldsymbol{1}}10$}
\psfrag{a8}[cb][cb][\scale]{${\boldsymbol{1}}11$}
\psfrag{b1}[cb][cb][\scale]{$0{\boldsymbol{0}}0$}
\psfrag{b2}[cb][cb][\scale]{$0{\boldsymbol{0}}1$}
\psfrag{b3}[cb][cb][\scale]{$0{\boldsymbol{1}}0$}
\psfrag{b4}[cb][cb][\scale]{$0{\boldsymbol{1}}1$}
\psfrag{b5}[cb][cb][\scale]{$1{\boldsymbol{0}}0$}
\psfrag{b6}[cb][cb][\scale]{$1{\boldsymbol{0}}1$}
\psfrag{b7}[cb][cb][\scale]{$1{\boldsymbol{1}}0$}
\psfrag{b8}[cb][cb][\scale]{$1{\boldsymbol{1}}1$}
\psfrag{c1}[cb][cb][\scale]{$00{\boldsymbol{0}}$}
\psfrag{c2}[cb][cb][\scale]{$00{\boldsymbol{1}}$}
\psfrag{c3}[cb][cb][\scale]{$01{\boldsymbol{0}}$}
\psfrag{c4}[cb][cb][\scale]{$01{\boldsymbol{1}}$}
\psfrag{c5}[cb][cb][\scale]{$10{\boldsymbol{0}}$}
\psfrag{c6}[cb][cb][\scale]{$10{\boldsymbol{1}}$}
\psfrag{c7}[cb][cb][\scale]{$11{\boldsymbol{0}}$}
\psfrag{c8}[cb][cb][\scale]{$11{\boldsymbol{1}}$}
\psfrag{x10}[cb][cb][\scale]{$\mcX_{1,0}$}
\psfrag{x11}[cb][cb][\scale]{$\mcX_{1,1}$}
\psfrag{x20}[cb][cb][\scale]{$\mcX_{2,0}$}
\psfrag{x21}[cb][cb][\scale]{$\mcX_{2,1}$}
\psfrag{x30}[cb][cb][\scale]{$\mcX_{3,0}$}
\psfrag{x31}[cb][cb][\scale]{$\mcX_{3,1}$}
\psfrag{a}[cb][cb][\scale]{$\Axkbkb{1,0}=6$, $\mc{I}_{\mc{X}_{1,0}}=\set{1,2,3,4}$}
\psfrag{b}[cb][cb][\scale]{$\Axkbkb{1,1}=6$, $\mc{I}_{\mc{X}_{1,1}}=\set{5,6,7,8}$}
\psfrag{c}[cb][cb][\scale]{$\Axkbkb{2,0}=4$, $\mc{I}_{\mc{X}_{2,0}}=\set{1,2,5,6}$}
\psfrag{d}[cb][cb][\scale]{$\Axkbkb{2,1}=4$, $\mc{I}_{\mc{X}_{2,1}}=\set{3,4,7,8}$}
\psfrag{e}[cb][cb][\scale]{$\Axkbkb{3,0}=0$, $\mc{I}_{\mc{X}_{3,0}}=\set{1,3,5,7}$}
\psfrag{f}[cb][cb][\scale]{$\Axkbkb{3,1}=0$, $\mc{I}_{\mc{X}_{3,1}}=\set{2,4,6,8}$}
\begin{center}
	\includegraphics{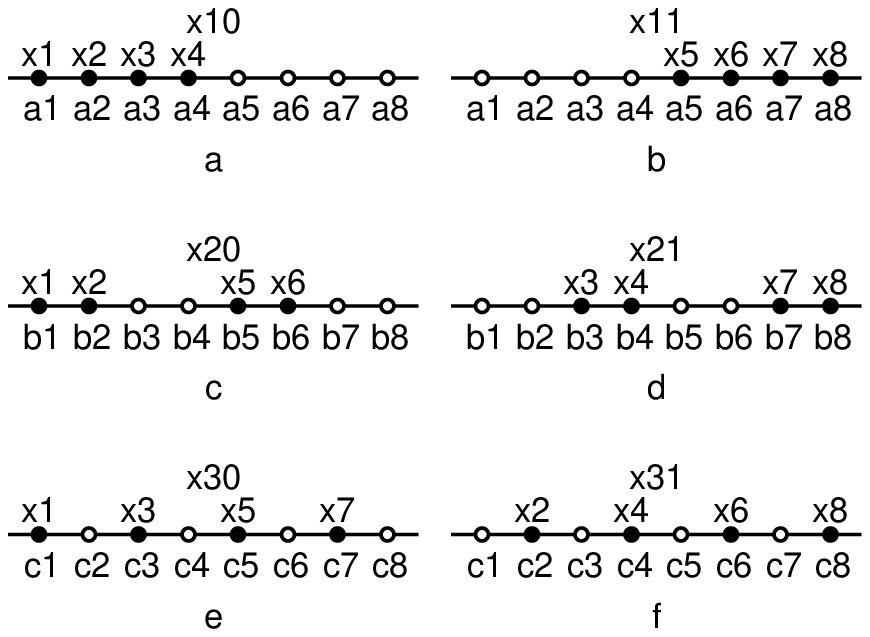}
	\caption[]{Subconstellations $\mcX_{k,b}$ (black circles) for $8$PAM labeled by the \gls{nbc} $\labeling=[0,1,2,3,4,5,6,7]$, where the values of $q_{i,k}$ for $k=1,2,3$ are highlighted in boldface, and where $\Ax=14$ and $\Cxl=22$. The values of $\Axkb$ and $\mcIXkb$ are also shown.}
	\label{8_pam_subconstellations}
\end{center}
\end{figure}
\begin{example}\label{8PAM.Ex}
In \figref{8_pam_subconstellations}, we show the $6$ subconstellations for an $8$PAM constellation labeled by the \gls{nbc} $\labeling=[0,1,2,3,4,5,6,7]$ \cite[Sec.~II-B]{Agrell10b}, as well as the corresponding values of $\mc{I}_{\mc{X}_{k,b}}$ and $\Axkb$ (defined below).
\end{example}

In BICM, the coded bits $\bQ=[Q_1,Q_2,\ld,Q_m]$ at the input of the mapper $\Phi$ (see \figref{BICM_Model}) are assumed to be independent but possibly nonuniformly distributed. Therefore, the vector of bit probabilities $[\pmf_{Q_1}(0),\pmf_{Q_2}(0),\ld,\pmf_{Q_m}(0)]$ induces a symbol input distribution $\PMF$ via the labeling as \cite[eq.~(8)]{Fabregas10a}, \cite[eq.~(9)]{Nguyen11}
\begin{align}\label{Pxi.BICM}
\pmf_{X}(x_i) =  \Pxi = \prod_{k=1}^{m} \pmf_{Q_k}(q_{i,k}).
\end{align}
Using \eqref{Pxi.BICM}, we obtain the conditional probabilities
\begin{align}\label{Pxi.Cku}
\pmf_{X|Q_k}(x|b) & =
\begin{cases}
\frac{\pmf_{X}(x)}{\pmf_{Q_k}(b)}, 	& \text{if $x\in\mcXkb$}\\
0, 	& \text{if $x\notin\mcXkb$}
\end{cases}
\end{align}
for $k=1,\ld,m$ and $b\in\set{0,1}$. According to \eqref{Pxi.Cku}, each of the $2m$ conditional input distributions $[\pmf_{X|Q_k}(x_1|b),\ld,\pmf_{X|Q_k}(x_M|b)]$ has $M/2$ non-zero probabilities, which specify which of the $M/2$ symbols in $\mcX$ are included in $\mcXkb$.
We shall use $X_{k,b}$ to denote a random variable with support $\mcXkb$ and \gls{pmf} $\pmf_{X|Q_k}(x|b)$. To shorten notation, we denote this \gls{pmf} by $\PMFkb$.

We next apply the results of \secref{Sec:Asymptotics} to BICM. To this end, we will often replace $\mcX$ and $\PMF$ in \secref{Sec:Asymptotics} by $\mcXkb$ and $\PMFkb$, respectively. Note, however, that $\MED$ as defined in \eqref{MED} still denotes the \gls{med} of the constellation $\mcX$. We will not consider the \gls{med} for subconstellations. This implies that it is possible that no pairs of constellation points in $\mcXkb$ are at \gls{med} (see, for example, $\mcX_{3,0}$ and $\mcX_{3,1}$ in \figref{8_pam_subconstellations}). It follows that the bounds \eqref{A.Bounds} on $\Ax$ modify to
\begin{equation}\label{A.Bounds.subconstellations}
0 \leq \Axkb \leq 2\left({M}/{2}-1\right).
\end{equation}

\subsection{Binary Labelings and Key Quantities for BICM}\label{binary.labelings}

The \gls{nbc} \cite[Sec.~II-B]{Agrell10b} is defined as the binary labeling $\labeling$ where $l_i=i-1$. It is the only optimal labeling for BICM in the low-\gls{snr} regime for $\mcX=\mcE$ \cite[Theorem~14]{Agrell10b}, \cite{Agrell12c}. A labeling $\labeling$ is said to be a \gls{gc} if for all $i,j$ for which $|x_i-x_j|=d$, the binary labels $\bq_i$ and $\bq_j$ are at Hamming distance one. One of the most popular \glspl{gc} is the \gls{brgc} \cite{Gray53,Agrell04,Agrell07}.

To characterize binary labelings we define the constant
\begin{align}
\label{C}
\Cxl 
& \triangleq\sumk\sum_{i\in\mcIXkz}\sum_{\substack{j\in\mcIXko\\|x_{i}-x_{j}|=\MED}} 2
\end{align}
which corresponds to twice the total number of \emph{different} bits between the labels of constellation symbol pairs at MED. For every given $x_{i}\in \mcXkz$, the inner sum in \eqref{C} considers all the constellation points in the subconstellation $\mcXko$ at \gls{med} from $x_i\in \mcXkz$. According to this interpretation, \eqref{C} can alternatively be expressed as
\begin{align}
\Cxl &= \sumk \left(\Ax-\Axkz-\Axko\right)\label{C.2}
\end{align}
where $\Ax-\Axkz-\Axko$ corresponds to twice the number of pairs of constellation points at \gls{med} with different labeling at bit position $k$. For example, for the constellation and labeling in \figref{8_pam_subconstellations}, $\Ax=14$ and $\Cxl=22$.

For $\mcX=\mcE$ and the \gls{nbc}, $\Cxl$ can be expressed as
\begin{align}
\nonumber
\CelNBC	& = 2\sumk (2^k-1) \\
\label{C_NBC}
					& = 2(2M-m-2)
\end{align}
which is obtained by noting that, for each $k$, there are $2^k-1$ symbols satisfying $q_{i,k}\neq q_{i+1,k}$, for $i=1,2,\dots,M-1$.

Note that, while $\Ax$ in \eqref{A} depends only on the geometry of the constellation, $\Cxl$ in \eqref{C} depends on both the geometry of the constellation and the labeling. Since any pair of constellation points at \gls{med} will differ in at least one bit, we have for any $\mcX$ and $\labeling$
\begin{align}\label{Cxl.Ax.Bound}
\Cxl\geq\Ax.
\end{align}

To state our main results on \gls{bicm}, and in analogy to \eqref{B}, we define the constant
\begin{align}
\label{D}
\Dxl 
& \triangleq \sumk\sum_{i\in\mcIXkz}\sum_{\substack{j\in\mcIXko\\|x_{i}-x_{j}|=\MED}} 2\sqrt{\Pxj\Pxi}.
\end{align}
Analog to \eqref{B.A.Uniform}, for a uniform input distribution
\begin{align}\label{E.C.uniform}
\Dxlu = \frac{\Cxl}{M}.
\end{align}

\subsection{Asymptotic Characterization of BICM}\label{Sec:BICM.Asymptotic}

The \gls{bicm-gmi} is an achievable rate for BICM \cite{Martinez09} and is one of the key quantities used to analyze BICM systems. For any $\PMF$ and $\labeling$, the \gls{bicm-gmi} is defined as \cite[eq.~(10)]{Martinez09}, \cite[eqs.~(32) and (41)]{Agrell10b}\footnote{Even though the \gls{bicm-gmi} is fully determined by the bit probabilities $[\pmf_{Q_1}(0),\pmf_{Q_2}(0),\ld,\pmf_{Q_m}(0)]$, we express it as a function of the input distribution $\PMF$ in \eqref{Pxi.BICM}.}
\begin{align}
\MIBIxp 	 & \triangleq \sumk I(Q_k;Y)\\
& = \sumk\Biggl(\MIxp-\sumb \pmf_{Q_k}(b)\MIxpkb\Biggr). \label{BICM.GMI.General}
\end{align}
Twice the derivative of $\MIBIxp$ is given by \cite[eq.~(3)]{Fabregas07}\footnote{Since the \gls{bicm-gmi} is not an \gls{mi}, its derivative is not an MMSE \cite{Fabregas07}. We thus avoid using the name MMSE, although we do use the MMSE-like notation $\MMSEBIxp$.}
\begin{align}
\label{BICM.GMMSE.General.0}
\MMSEBIxp 	 & \triangleq 2\frac{\tr{d}\MIBIxp}{\tr{d}\rho}\\
			 & = \sumk\Biggl(\MMSExp-\sumb \pmf_{Q_k}(b)\MMSEkpkb\Biggr) \label{BICM.GMMSE.General}.
\end{align}
In these expressions, $\MIxpkb$ and $\MMSEkpkb$ are defined, in analogy to \eqref{mi-def}--\eqref{mmse-def}, as
\begin{equation}
\MIxpkb \triangleq \Ex_{X_{k,b},Y}\left[\log\left({\pdf_{Y|X_{k,b}}(Y|X_{k,b})}/{\pdf_{Y}(Y)}\right)\right]
\end{equation}
and
\begin{align}
\MMSEkpkb &\triangleq \Ex_{X_{k,b},Y}[(X_{k,b}-\hat{X}_{\PMFkb}^{\textME}(Y))^2]\\
\hat{X}_{\PMFkb}^{\textME}(y) &\triangleq \Ex_{X_{k,b}}[X_{k,b}|Y=y]
\end{align}
where $Y$ is the random variable resulting from transmitting $X_{k,b} \in \mcXkb$ over the \gls{awgn} channel \eqref{AWGN}.

We define the \gls{bep} as\footnote{Note that \eqref{bep} is the \gls{bep} averaged over the $m$ bit positions, in contrast to the \gls{bicm-gmi} in \eqref{BICM.GMI.General}, which is a sum of $m$ bit-wise \glspl{mi}.}
\begin{align}\label{bep}
\BEPxp \triangleq \frac{1}{m}\sumk \Pr\set{\hat{Q}_k^{\textMAP}(Y)\ne Q_k}
\end{align}
where $Q_k$ is the transmitted bit and $\hat{Q}_k^{\textMAP}(Y)$ is a hard-decision on the bit, \ie $[\hat{Q}_1^{\textMAP}(y),\ld,\hat{Q}_m^{\textMAP}(y)]=\Phi^{-1}(\hat{X}^{\textMAP}(y))$ with $\hat{X}^{\textMAP}(y)$ given by \eqref{map}.\footnote{The \gls{bep} in \eqref{bep} is based on hard-decisions made by the \emph{symbol-wise} \gls{map} demapper. Alternatively, one could study a \emph{bit-wise} \gls{map} demapper for which $\hat{Q}_k^{\textMAP}(y) = \argmax_{b\in\set{0,1}}{\pmf_{Q_k|Y}(b|y)}$. This demapper minimizes the \gls{bep} \cite{Ivanov13a} (see also \cite{Ivanov12a}), but its analysis is much more involved.}

The \gls{bicm-gmi} tends to $\ENX$ as $\rho$ tends to infinity. The following theorem shows how fast $\MIBIxp$ converges to $\ENX$.
\begin{theorem}\label{BICM-GMI.Asym.Theo}
For any $\PMF$ and $\labeling$
\begin{align}
\limrinf{
\frac{\ENX-\MIBIxp}{\QF\left({\sqrt{\rho}\MED}/{2}\right)}
} & = \pi \Dxl \label{BICM-GMI.Asymptotic}
\end{align}
where $\Dxl$ is given by \eqref{D}.
\end{theorem}
\begin{IEEEproof}
See \appref{BICM-GMI.Asym.Proof}.
\end{IEEEproof}

Similar to Theorems \ref{MMSE.Gen.Asym.Theo} and \ref{SEP.Gen.Asym.Theo}, we have following asymptotic expressions for $\MMSEBIxp$ and the \gls{bep}.
\begin{theorem}\label{BICM-GMMSE.Asym.Theo}
For any $\PMF$ and $\labeling$
\begin{align}
\limrinf{
\frac{\MMSEBIxp}{\QF\left({\sqrt{\rho}\MED}/{2}\right)}
}  & = \frac{\pi\MED^2}{4} \Dxl. \label{BICM-GMMSE.Asymptotic}
\end{align}
\end{theorem}
\begin{IEEEproof}
See \appref{BICM-GMMSE.Asym.Theo.Proof}.
\end{IEEEproof}
\begin{theorem}\label{BEP.Gen.Asym.Theo}
For any $\PMF$ and $\labeling$
\begin{align}\label{BEP.Gen.Asymptotic}
\limrinf{
\frac{\BEPxp}{\QF\left({\sqrt{\rho}\MED}/{2}\right)}
}
= \frac{\Dxl}{m}.
\end{align}
\end{theorem}
\begin{IEEEproof}
See \appref{BEP.Gen.Asym.Theo.Proof}.
\end{IEEEproof}

It follows from Theorems~\ref{BICM-GMI.Asym.Theo}--\ref{BEP.Gen.Asym.Theo} that, at high \gls{snr}, the \gls{bicm-gmi}, $\MMSEBIxp$, and the \gls{bep} behave as
\begin{align}
\label{BICM-GMI.approx}
\MIBIxp &\approx \ENX - \pi \Dxl\QF\left({\sqrt{\rho}\MED}/{2}\right)\\
\label{BICM-GMMSE.approx}
\MMSEBIxp &\approx  \frac{\pi\MED^2}{4} \Dxl \QF\left({\sqrt{\rho}\MED}/{2}\right)\\
\label{BEP.approx}
\BEPxp &\approx \frac{\Dxl}{m}\QF\left({\sqrt{\rho}\MED}/{2}\right).
\end{align}

For a given constellation and input distribution, the results in \eqref{BICM-GMI.approx}--\eqref{BEP.approx} indicate that, at high \gls{snr}, a maximization of the \gls{bicm-gmi} over binary labelings is equivalent to a minimization of both its derivative and the \gls{bep}.

\begin{table}[t]
\renewcommand{\arraystretch}{1.5}
\caption{Different parameters for the constellation and input distributions in Example \ref{NES.4PAM.BICM.Ex}.}\label{Table.Ex}
\centering
\small
\begin{tabular}{c|c|c|c|c}
\hline

\hline
$\bp$ & $\pmf_{Q_1}(0), \pmf_{Q_2}(0)$ & $\ENX$ & $\Dxl$ & $\MED$ \\
\hline
\hline
$\bp'$   & $1/2, 1/2$ & $1.3863$ & $1.0000$ & $0.6325$ \\
\hline
$\bp''$  & $1/2, 1/4$ & $1.2555$ & $0.8660$ & $0.7559$ \\
\hline
$\bp'''$ & $4/5, 4/5$ & $1.0008$ & $0.8000$ & $0.5423$ \\
\hline

\hline
\end{tabular}
\end{table}
\begin{figure}[tpb]
\newcommand{\scale}{0.8}
\centering
\psfrag{xlabel}[cc][cB][\scale]{$\rho$~[dB]}%
\psfrag{ylabel}[cb][ct][\scale]{$\MIBIxp$~[nats/symbol]}%
\psfrag{PU}[cl][cl][\scale]{$\bp'$}%
\psfrag{P1}[cl][cl][\scale]{$\bp''$}%
\psfrag{P2}[cl][cl][\scale]{$\bp'''$}%
\includegraphics[width=0.97\columnwidth]{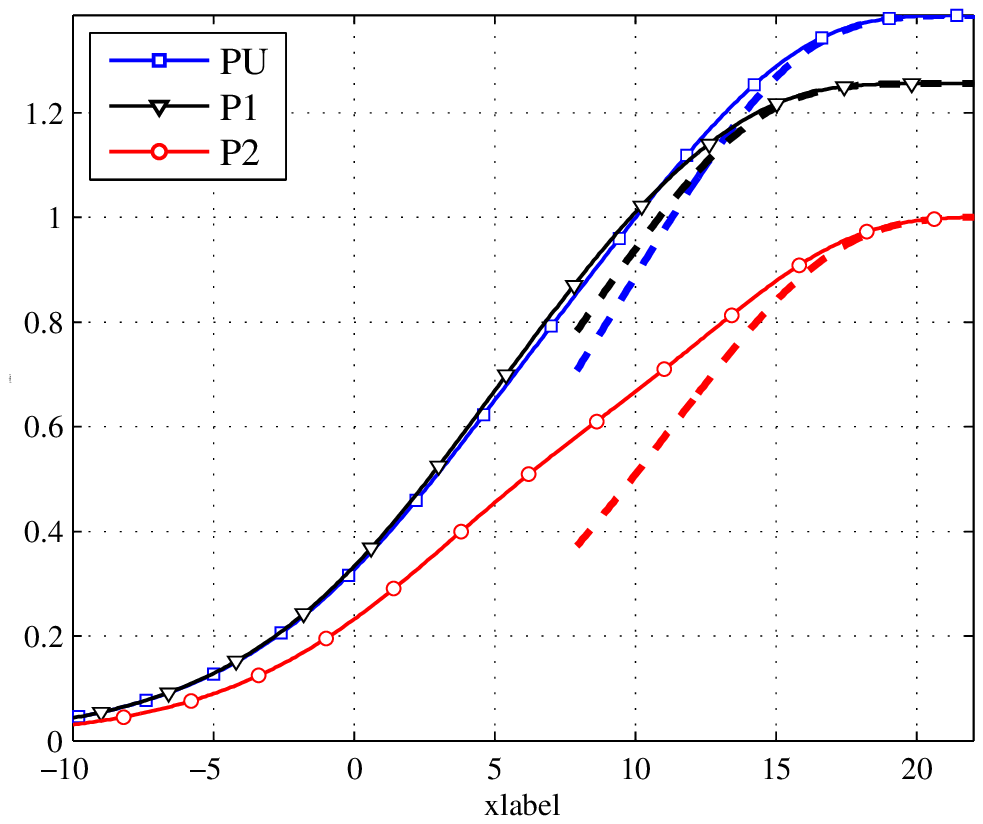}
\caption{$\MIBIxp$ for the three input distributions in \exref{NES.4PAM.BICM.Ex} and the constellation $\mcX=\set{\pm 4,\pm 2}$ (normalized to $\Es=1$) (solid lines with markers) and the asymptotic expression in \eqref{BICM-GMI.approx} (thick dashed lines).}
\label{MutualInformation_NES_4PAM}
\end{figure}
\begin{figure}[tpb]
\newcommand{\scale}{0.8}
\centering
\psfrag{xlabel}[cc][cB][\scale]{$\rho$~[dB]}%
\psfrag{ylabel}[cb][ct][\scale]{$\ENX-\MIBIxp$~[nats/symbol]}%
\psfrag{PU}[cl][cl][\scale]{$\bp'$}%
\psfrag{P1}[cl][cl][\scale]{$\bp''$}%
\psfrag{P2}[cl][cl][\scale]{$\bp'''$}%
\includegraphics[width=0.97\columnwidth]{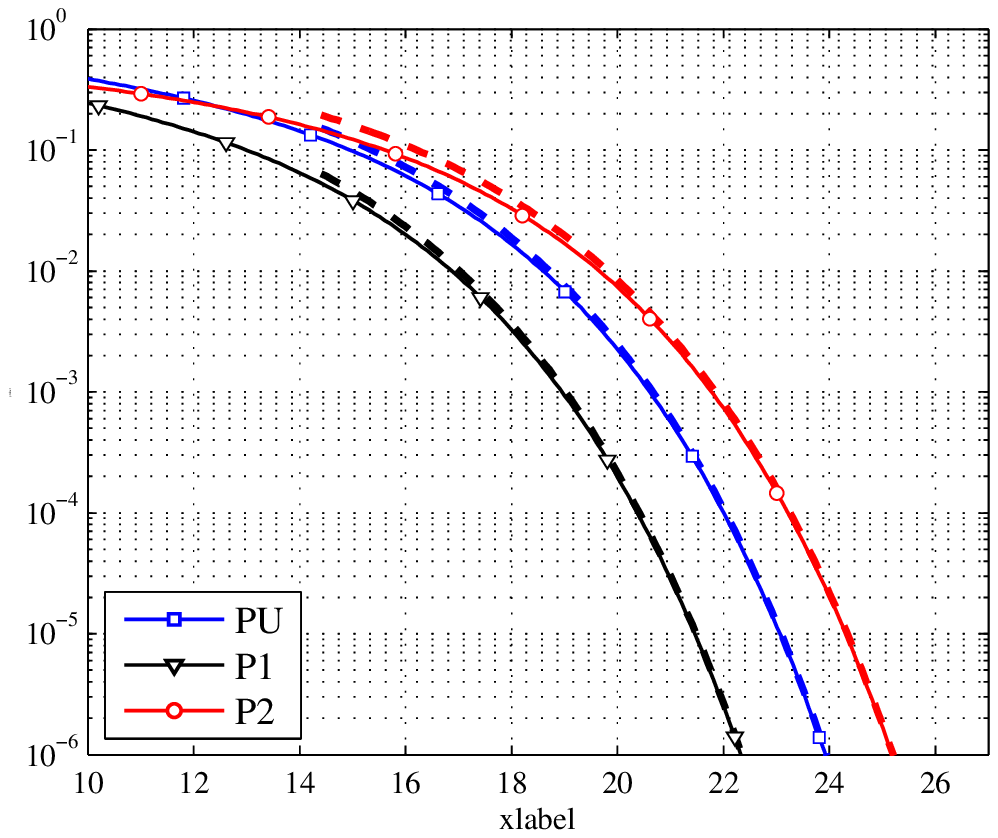}
\caption{$\ENX-\MIBIxp$ for the three input distributions in \exref{NES.4PAM.BICM.Ex} and the constellation $\mcX=\set{\pm 4,\pm 2}$ (normalized to $\Es=1$) (solid lines with markers) and the asymptotic expression in \eqref{BICM-GMI.approx} (thick dashed lines), \ie $\ENX - \MIBIxp \approx \pi \Dxl \QF\left({\sqrt{\rho}\MED}/{2}\right)$.}
\label{ConditionalEntropy_NES_4PAM}
\end{figure}
\begin{example}\label{NES.4PAM.BICM.Ex}
Consider the constellation $\mcX=\set{\pm 4,\pm 2}$ in \exref{NES.4PAM.Ex}
and the labeling $\labeling_\tr{GC}=[0,1,3,2]$, which gives $\Ax=\Cxl=4$. Furthermore, consider the three input distributions
\begin{align*}
\bp'  &=[1/4,1/4,1/4,1/4]\\
\bp'' &=[1/8,3/8,3/8,1/8]\\
\bp'''&=[16/25,4/25,1/25,4/25]
\end{align*}
which are induced by the bit probabilities listed in the second column of \tabref{Table.Ex}. \tabref{Table.Ex} further lists $\ENX$, $\Dxl$, and $\MED$ when the constellation is normalized to $\Es=1$. \figref{MutualInformation_NES_4PAM} shows the \gls{bicm-gmi} curves and \figref{ConditionalEntropy_NES_4PAM} shows the corresponding curves for $\ENX-\MIBIxp$. The asymptotic expression \eqref{BICM-GMI.approx} is also shown. Observe how this asymptotic expression approximates well the \gls{bicm-gmi} for a large range of \gls{snr}.
\end{example}

For a uniform input distribution, Theorems~\ref{BICM-GMI.Asym.Theo}--\ref{BEP.Gen.Asym.Theo} particularize to the following result.
\begin{corollary}\label{BICM.Uniform.Asym.Coro}
For any $\mcX$ and $\labeling$ and a uniform input distribution
\begin{align}
\label{BICM-GMI.Uniform.Asymptotic}
\limrinf{
\frac{\log{M}-\MIBIxu}{\QF\left({\sqrt{\rho}\MED}/{2}\right)}
}
&= \pi\frac{\Cxl}{M}\\
\label{BICM-GMMSE.Uniform.Asymptotic}
\limrinf{
\frac{\MMSEBIxu}{\QF\left({\sqrt{\rho}\MED}/{2}\right)}
}
&= \frac{\pi\MED^2}{4} \frac{\Cxl}{M}\\
\label{BEP.Uniform}
\limrinf{
\frac{\BEPxu}{\QF\left({\sqrt{\rho}\MED}/{2}\right)}
}
&= \frac{\Cxl}{mM}
\end{align}
where $\Cxl$ is given by \eqref{C}.
\end{corollary}
\begin{IEEEproof}
From Theorems~\ref{BICM-GMI.Asym.Theo}--\ref{BEP.Gen.Asym.Theo} and \eqref{E.C.uniform}.
\end{IEEEproof}

The expression in \eqref{BEP.Uniform} for the \gls{bep} is well-known, see, e.g., \cite[p.~130]{Madhow08_Book}. The asymptotic results for BICM are summarized in \tabref{Table.Summary.BI}.

\begin{table}[t]
\renewcommand{\arraystretch}{2.3}
\caption[]{Summary of asymptotics of the \gls{bicm-gmi}, twice its derivative, and the \gls{bep}.}\label{Table.Summary.BI}
\centering
\small
\begin{tabular}{c|cc}
\hline

\hline
Input Distribution 				& $\PMF$						& $\PMFU$					\\
\hline
\hline
$\underset{\rho\rightarrow\infty}{\lim}{\dfrac{\ENX-\MIBIxp}{\QF\left({\sqrt{\rho}\MED}/{2}\right)}
}$		& $\pi \Dxl$	& $\pi\dfrac{\Cxl}{M}$ \\
\hline
$\underset{\rho\rightarrow\infty}{\lim}{\dfrac{\MMSEBIxp}{\QF\left({\sqrt{\rho}\MED}/{2}\right)}
}$		& $\dfrac{\pi\MED^2}{4} \Dxl$ & $\dfrac{\pi\MED^2}{4} \dfrac{\Cxl}{M}$ \\[1.5ex]
\hline
$\underset{\rho\rightarrow\infty}{\lim}{
\dfrac{\BEPxp}{\QF\left({\sqrt{\rho}\MED}/{2}\right)}
}$		& $\dfrac{\Dxl}{m}$ & $\dfrac{\Cxl}{mM}$ \\[1.5ex]
\hline

\hline
\end{tabular}
\end{table}

\subsection{Classification of Labelings at high SNR}

To study the asymptotic behavior of the \gls{bicm-gmi} for different labelings $\labeling$, we introduce the two functions
\begin{align}
\label{KIxpl.Def}
\KMIxpl	&\triangleq \frac{\ENX-\MIBIxp}{\ENX-\MIxp}\\
\KMMSExpl	&\triangleq \frac{\MMSEBIxp}{\MMSExp}.
\label{KMMSExpl.Def}
\end{align}
Noting that $\MIBIxp \leq \MIxp$ \cite[eq.~(16)]{Caire98}, \cite[Theorem~5]{Agrell10b}, we have
\begin{equation}
\label{KIxpl.geq}
\KMIxpl \geq 1.
\end{equation}
We further define
\begin{align}
\label{Rxpl.Def}
\Rxpl 	& \triangleq \limrinf{\KMIxpl}\\
\label{Rxpl.MMSE}
  & = \limrinf{\KMMSExpl}
\end{align}
where \eqref{Rxpl.MMSE} follows from L'H\^{o}pital's rule.
Theorems~\ref{MI.Gen.Asym.Theo} and \ref{BICM-GMI.Asym.Theo} yield
\begin{align}\label{Rxpl.Final}
\Rxpl 	= \frac{\Dxl}{\Bxp}.
\end{align}
Furthermore, by \eqref{KIxpl.geq},
\begin{equation} \label{R.lowerbound}
\Rxpl\geq 1.
\end{equation}

We next study $\Rxpl$ for a uniform input distribution $\PMFU$. With a slight abuse of notation, we will refer to $\R_{\PMFU,\labeling}$ as $\Rxul$.

\begin{corollary} \label{RC.relation}
For any labeling $\labeling$ and constellation $\mcX$,
\begin{align}\label{Rxpl.Final.Uniform}
\Rxul 	= \frac{\Cxl}{\Ax}.
\end{align}
\end{corollary}
\begin{IEEEproof}
Follows by using \eqref{E.C.uniform} and \eqref{B.A.Uniform} in \eqref{Rxpl.Final}.
\end{IEEEproof}

By \cororef{RC.relation}, an upper bound on $\Cxl$ yields an upper bound on $\Rxul$.
\begin{theorem}\label{Bounds.Cxl.Theo}
For any one-dimensional constellation $\mcX$ and any labeling $\labeling$
\begin{align}\label{Bounds.Cxul}
\Cxl \leq \min \left( m\Ax, (m-1)\Ax+M \right)
\end{align}
and hence
\begin{align}\label{Bounds.Rxul}
\Rxul \leq
\frac{\min \left( m\Ax, (m-1)\Ax+M \right)}{\Ax}.
\end{align}
\end{theorem}
\begin{IEEEproof}
By definition, we have $\Axkz\geq 0$ and $\Axko\geq 0$ which by \eqref{C.2} yields
\begin{equation}
\label{Tobi_Bla}
\Cxl \le m\Ax.
\end{equation}
This bound holds with equality if the labels of all $\Ax/2$ pairs of constellation points at \gls{med} differ in exactly $m$ bits. Conversely, \eqref{Tobi_Bla} can only hold with equality if $\Ax\leq M$, since there are only $M/2$ pairs of labels at Hamming distance $m$. For $\Ax>M$, the quantity $\Cxl$ is maximized if the labels of $M/2$ constellation pairs differ in $m$ bits and the labels of the remaining $(\Ax-M)/2$ pairs differ in $m-1$ bits, which gives
\begin{align}
\Cxl & \le mM + (m-1)(\Ax-M) \\ & =(m-1)\Ax+M, \quad \Ax>M.\label{Tobi_BlaBla}
\end{align}
Combining \eqref{Tobi_Bla} and \eqref{Tobi_BlaBla} proves \eqref{Bounds.Cxul}, which together with \eqref{Rxpl.Final.Uniform} proves \eqref{Bounds.Rxul}.
\end{IEEEproof}

For an $M$PAM constellation, \theoref{Bounds.Cxl.Theo} specializes to
\begin{align}
\label{Bounds.Cel}
\Cel & \leq 2mM-2m-M+2\\
\label{Bounds.Reul}
\Reul &\leq m-\frac{M-2}{2M-2}.
\end{align}
Furthermore, if the $M$PAM constellation is labeled with the \gls{nbc}, we obtain from \eqref{C_NBC}
\begin{align}\label{R_NBC}
\ReulNBC&=\frac{2M-m-2}{M-1}.
\end{align}

\begin{figure}[t]
\centering
\newcommand{\scale}{0.75}
\psfrag{xlabel}[cc][cB][\scale]{$\rho$~[dB]}%
\psfrag{ylabel}[cc][ct][\scale]{}
\psfrag{MIL}[cl][cl][\scale]{$\KMIeul$}%
\psfrag{MMSEL}[cl][cl][\scale]{$\KMMSEeul$}%
\psfrag{H0}[tl][tl][\scale]{\hspace{-5mm}$\ReulBRGC=3/3$}%
\psfrag{H1}[tl][tl][\scale]{\hspace{-5mm}$\ReulNBC=4/3$}%
\psfrag{H2}[tl][tl][\scale]{\hspace{-5mm}$\ReulAGC=5/3$}%
\includegraphics[width=0.97\columnwidth]{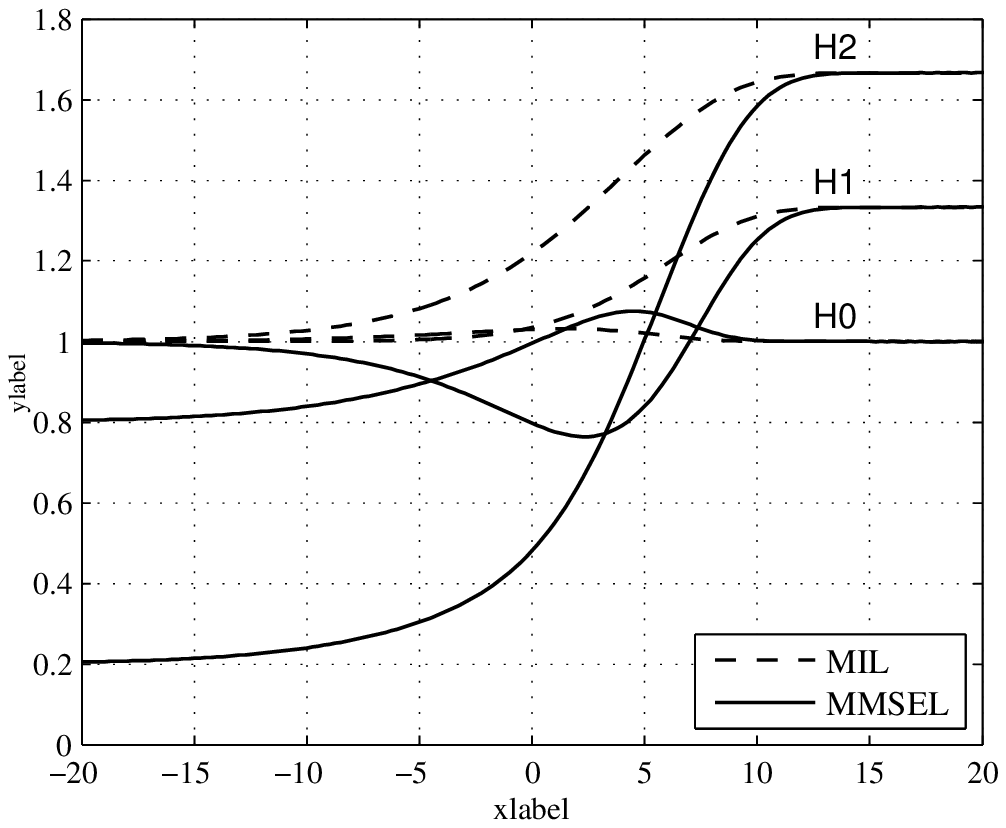}
\caption{Functions $\KMIeul$ (dashed lines) and $\KMMSEeul$ (solid lines) for $4$PAM (normalized to $\Es=1$) and all three nonequivalent labelings. The values of $\Reul$ are also shown.}
\label{MI_and_MMSE_ratios_4PAM_all_labelings_rhodB}
\end{figure}
\begin{example}\label{Example.4PAM}
In \figref{MI_and_MMSE_ratios_4PAM_all_labelings_rhodB}, we show the functions $\KMIeul$ and $\KMMSEeul$ in \eqref{KIxpl.Def} and \eqref{KMMSExpl.Def}, respectively, for a $4$PAM constellation with a uniform input distribution ($\PMF=\PMFEU$, $\Ax=6$) and the three labelings that give a different BICM-GMI: $\labeling_\tr{GC}=[0,1,3,2]$, $\labeling_\tr{NBC}=[0,1,2,3]$, and $\labeling_\tr{AGC}=[0,3,2,1]$.\footnote{The \gls{agc} will be formally introduced in \secref{GCs.AGCs}.} The values of $\Reul$ are also shown. In contrast to the \gls{bicm-gmi} curves plotted, \eg in \cite[Fig.~3]{Stierstorfer07a} and \cite[Fig.~1]{Fabregas07}, the functions $\KMIeul$ and $\KMMSEeul$ allow us to study different labelings at high \gls{snr}. Observe that the \gls{gc} gives $\ReulBRGC=1$, and that the \gls{agc} achieves the upper bound in \eqref{Bounds.Reul}, \ie $\ReulAGC=5/3$.
The function $\KMMSEeul$ also allows us to study different labelings at low \gls{snr}: \figref{MI_and_MMSE_ratios_4PAM_all_labelings_rhodB} shows that the \gls{nbc} is the binary labeling for $4$PAM that gives the largest value for $\MMSEBIxu$ as $\rho$ tends to zero, which agrees with \cite{Stierstorfer09a}, \cite[Theorem~14]{Agrell10b}.\footnote{The relationship between the coefficient $\alpha$ determining the low-\gls{snr} behavior of a zero-mean constellation with a uniform input distribution \cite[eq.~(47)]{Agrell10b} and $\KMMSEeul$ is $\alpha\log{2}=\lim_{\rho\rightarrow 0}\KMMSEeul$ (see also \cite[eq.~(86)]{Guo05}).} 
\end{example}

\begin{figure}[tb]
\newcommand{\scale}{0.89}
\newcommand{\ww}{1}
\centering
\psfrag{H0}[cc][cc][\scale]{$\ReulBRGC=\frac{7}{7}$}%
\psfrag{H1}[cl][cl][\scale]{}
\psfrag{H2}[cl][cl][\scale]{}
\psfrag{H3}[cl][cl][\scale]{}
\psfrag{H4}[cc][cc][\scale]{$\ReulNBC=\frac{11}{7}$}%
\psfrag{H5}[cl][cl][\scale]{}
\psfrag{H6}[cl][cl][\scale]{}
\psfrag{H7}[cl][cl][\scale]{}
\psfrag{H8}[cl][cl][\scale]{}
\psfrag{H9}[cl][cl][\scale]{}
\psfrag{H10}[cl][cl][\scale]{}
\psfrag{H11}[cc][cc][\scale]{$\ReulAGC=\frac{18}{7}$}%
\psfrag{xlabel}[cc][cB][\scale]{$\rho$~[dB]}%
\psfrag{ylabel}[cc][ct][\scale]{$\KMMSEeul$}%
\includegraphics[width=0.97\columnwidth]{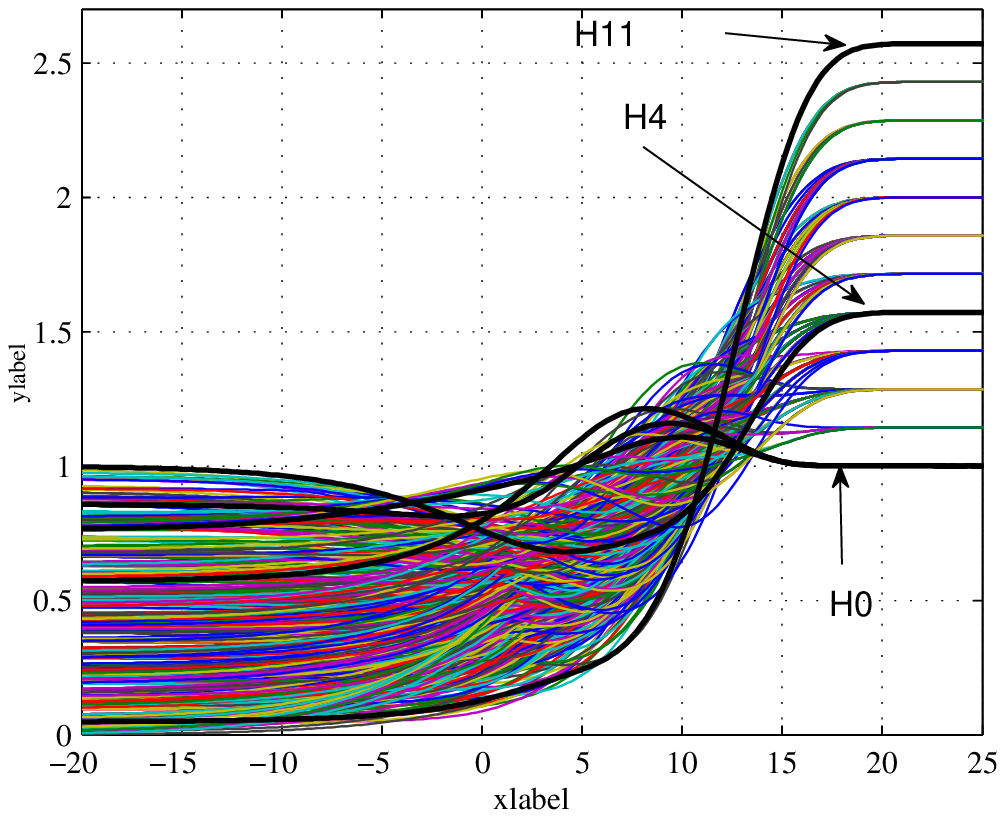}
\caption[]{$\KMMSEeul$ for the $458$ labelings that give a different \gls{bicm-gmi} for $8$PAM (normalized to $\Es=1$). The values of $\Reul$ for the three nonequivalent \glspl{gc}, the \gls{nbc}, and the AGC are also shown (thick lines).}
\label{MMSE_ratios_8PAM_all_labelings_rhodB}
\end{figure}
\begin{example}\label{Example.8PAM.v3}
In \figref{MMSE_ratios_8PAM_all_labelings_rhodB}, we show the function $\KMMSEeul$ for $8$PAM ($\PMF=\PMFEU$, $\Ax=14$) and all the $458$ labelings that give a different BICM-GMI \cite{Alvarado11b}. The value $\ReulNBC$ obtained using \eqref{R_NBC} is also shown. We further highlight the three nonequivalent \glspl{gc} (in terms of \gls{bep}) \cite[Table~I]{Agrell04}: the \gls{brgc} $\labeling=[0,1,3,2,6,7,5,4]$, $\labeling=[0,1,3,2,6,4,5,7]$, and $\labeling=[0,1,3,7,5,4,6,2]$. All these \glspl{gc} give $\Reul=1$. Observe that there are $12$ possible values of $\Reul$, which is consistent with \cite[Fig.~3]{Alvarado11b}.\footnote{Further note that $\lim_{\rho\rightarrow 0}\KMMSEeul$ reveals the 72 classes of labelings reported in \cite[Fig.~6~(a)]{Agrell10b}.} Using \eqref{BEP.Uniform}, the $12$ values of $\Reul$ in \figref{MMSE_ratios_8PAM_all_labelings_rhodB} translate into $12$ different asymptotic \gls{bep} curves, which were recently reported in \cite[Fig.~4]{Ivanov12a}.
\end{example}

\begin{figure}[t]
\newcommand{\scale}{0.89}
\newcommand{\ww}{1}
\centering
\psfrag{UB}[cc][cc][\scale]{Upper Bound}%
\psfrag{NBC}[bc][bc][\scale]{NBC}%
\psfrag{BRGC}[cc][cc][\scale]{GCs}%
\psfrag{xlabel}[cc][cB][\scale]{$t$}%
\psfrag{ylabel}[cc][ct][\scale]{$O_{\Rxul}(t)$}%
\includegraphics[width=0.97\columnwidth]{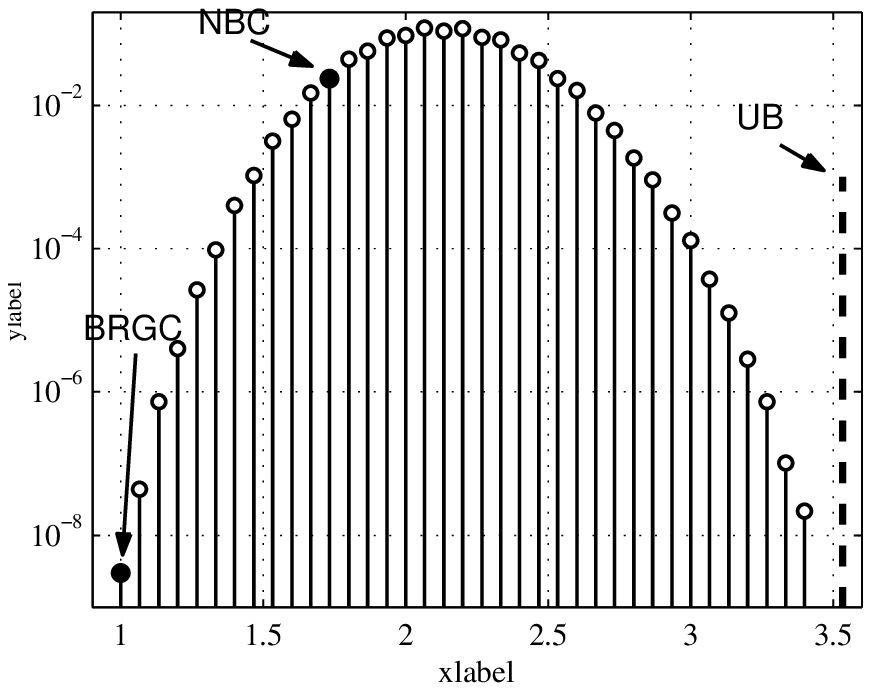}
\caption[]{Approximated $O_{\Rxul}(t)$ using $10^{9}$ randomly generated labelings for $16$PAM (normalized to $\Es=1$). For \glspl{gc}, we have $\ReulBRGC=1$ and for the \gls{nbc}, we have $\ReulNBC=26/15$. The upper bound \eqref{Bounds.Reul} is also shown.}
\label{H_1000000000_labelings_M_16_log}
\end{figure}
\begin{example}
\label{Example7}
Motivated by \cite[Fig.~6]{Agrell10b}, we study the relative occurrence of labelings with a given $\Rxul$, \ie
\begin{equation}
O_{\Rxul}(t) \triangleq \frac{\textnormal{number of labelings with $\Rxul=t$}}{\textnormal{total number of labelings}}.
\end{equation}
In \figref{H_1000000000_labelings_M_16_log}, we present an approximation of $O_{\Rxul}(t)$ for $16$PAM, obtained by randomly generating $10^{9}$ labelings. We highlight $\ReulBRGC$ and $\ReulNBC$. The upper bound \eqref{Bounds.Reul} is also shown. Observe that most of the possible labelings are not Gray.
\end{example}

\subsection{Gray Codes, Anti-Gray Codes, and Asymptotic Optimality}\label{GCs.AGCs}

In view of the lower bound \eqref{R.lowerbound}, we say that a labeling $\labeling$ is \gls{ao} in terms of \gls{bicm-gmi} for a constellation $\mcX$ and a uniform input distribution if it satisfies $\Rxul=1$. Intuitively, an \gls{ao} labeling is a binary labeling for which the \gls{bicm-gmi} approaches $\ENX$ as fast as the \gls{mi} does for the same constellation $\mcX$.

By inspection of \eqref{R_NBC}, we see that the \gls{nbc} for $M$PAM is not an \gls{ao} labeling for $m\ge 2$. The following theorem demonstrates that \glspl{gc} are \gls{ao} at high \gls{snr}. Thus, it proves the conjecture of the optimality of \glspl{gc} at high SNR in terms of \gls{bicm-gmi} \cite[Sec.~III-C]{Caire98}.
\begin{theorem}\label{Gray.Optimal.Theo}
For any constellation $\mcX$ and a uniform input distribution, a labeling is \gls{ao} if and only if it is a \gls{gc}.
\end{theorem}
\begin{IEEEproof}
By definition, for a \gls{gc}, all pairs of labelings of constellation points at \gls{med} are at Hamming distance one. Thus, $\Ax=\Cxl$, and by \eqref{Rxpl.Final.Uniform}, $\Rxul=1$, demonstrating that every \gls{gc} is \gls{ao}. Conversely, for every non-\gls{gc}, there is at least one pair of constellation points at \gls{med} with Hamming distance larger than one. Consequently, every non-\gls{gc} gives $\Cxl > \Ax$, and therefore, $\Rxul>1$.
\end{IEEEproof}

\begin{remark}
The optimality of \glspl{gc} directly extends to multidimensional constellations that are constructed as direct products of one-dimensional constellations, provided that the labeling is generated via an ordered direct product of \glspl{gc}. This construction of constellation and labelings was formally studied \eg in \cite[Theorem~15]{Agrell10b}.
\end{remark}

\begin{remark}
While the \gls{nbc} is not \gls{ao} for an $M$PAM constellation, it may be \gls{ao} for an unequally spaced constellation. For example, this is the case if the \gls{nbc} is used with the constellation in Example~\ref{NES.4PAM.Ex}, in which case the \gls{nbc} is a \gls{gc}.
\end{remark}

\theoref{Gray.Optimal.Theo} shows that \glspl{gc} minimize $\Rxul$. In what follows, we show that, for $M$PAM constellations, it is always possible to construct a labeling that maximizes $\Reul$, \ie a labeling that achieves the upper bound \eqref{Bounds.Reul}.

Let $\setCx$ denote the set of all possible values that $\Cxl$ can take.  Noting that $\Cxl$ is an even integer bounded by \eqref{Cxl.Ax.Bound} and \eqref{Bounds.Cxul}, it follows that, for any constellation $\mcX$, the cardinality of $\setCx$ satisfies
\begin{align}\label{Size.Cxl}
|\setCx|\le
\frac{1}{2}\min \left\{ (m-1)\Ax+2, (m-2)\Ax+M+2 \right\}.
\end{align}
The expression \eqref{Size.Cxl} is an upper bound on the number of classes of labelings with different high-\gls{snr} behavior in terms of \gls{bicm-gmi} (or equivalently \gls{bep}). For the particular case of $\mcX=\mcE$, we obtain from \eqref{A.Uniform} and \eqref{Size.Cxl}
\begin{align}\label{Size.Cel}
|\setCe|\le mM-\frac{3M}{2}-m+3.
\end{align}
For example, for $4$PAM we have $|\setCe| \le 3$ and for $8$PAM we have $|\setCe| \le 12$, which is consistent with the $3$ and $12$ classes at high \gls{snr} shown in \figref{MI_and_MMSE_ratios_4PAM_all_labelings_rhodB} and \figref{MMSE_ratios_8PAM_all_labelings_rhodB}, respectively. For $16$PAM, the upper bound \eqref{Size.Cel} indicates that there are at most $39$ classes. However, \figref{H_1000000000_labelings_M_16_log} shows only $37$ classes, all giving rise to a $\Reul$ strictly smaller than \eqref{Bounds.Reul}. This raises the question of whether to produce \figref{H_1000000000_labelings_M_16_log} we were drawing not enough labelings\footnote{Without discarding trivial operations, there are $16! \approx 2.1 \cd 10^{13}$ labelings, so randomly generating $10^9$ labelings covers only a small fraction of all possible labelings.} or whether the upper bounds \eqref{Bounds.Reul} and \eqref{Size.Cel} are loose. As we shall show next, \eqref{Bounds.Reul} is achieved by an \gls{agc}.

The \gls{agc} of order $m\ge 2$ is defined by the $M\times m$ binary matrix $\AGC_{m}$ (the $i$th row is the binary label for $x_i$) where $\AGC_{m}$ is constructed according to the following recursive procedure:

Let $\AGC_{1}=[0,1]\T$. Construct $\AGC_m$ from $\AGC_{m-1}$ following the next three steps:
\begin{enumerate}[Step 1]
\item Reverse the order of the $M/2$ rows in $\AGC_{m-1}$, and append them below $\AGC_{m-1}$ to construct a new matrix $\AGC'_{m}$ with $M$ rows and $m-1$ columns.
\item Append the length $M$ column vector $[0,1,0,1,\ld,0,1]\T$ to the left of $\AGC'_{m}$ to create $\AGC''_{m}$, with $M$ rows and $m$ columns.
\item Negate all bits in the lower half of $\AGC''_{m}$ to obtain $\AGC_{m}$.
\end{enumerate}

\begin{figure*}[t]
\centering
\newcommand{\scale}{0.85}
\psfrag{0}[cc][cc][\scale]{0}
\psfrag{1}[cc][cc][\scale]{1}
\psfrag{A1}[cc][cc][\scale]{$\AGC_1$}
\psfrag{A2p}[bc][Bc][\scale]{$\AGC_2'$}
\psfrag{A2pp}[cc][cc][\scale]{}
\psfrag{A2ppp}[cc][cc][1.3]{$\overbrace{~~~~~~~~}^{\AGC_2''}$}
\psfrag{A2}[cc][cc][\scale]{$\AGC_2$}
\psfrag{A3p}[bc][Bc][\scale]{$\AGC_3'$}
\psfrag{A3pp}[cc][cc][\scale]{}
\psfrag{A3ppp}[cc][cc][1.2]{$\overbrace{~~~~~~~~~~~}^{\AGC_3''}$}
\psfrag{A3}[cc][cc][\scale]{$\AGC_3$}
\psfrag{m2}[cr][cr][\scale]{$m=2$}
\psfrag{m3}[cr][cr][\scale]{$m=3$}
\psfrag{append}[cl][cl][\scale]{\emph{append}}
\psfrag{repeat}[cr][cr][\scale]{\emph{repeat}}
\psfrag{negate}[cr][cr][\scale]{\emph{negate}}
\psfrag{reflect}[cl][cl][\scale]{\emph{reverse}}
\begin{center}
	\includegraphics[width=0.65\textwidth]{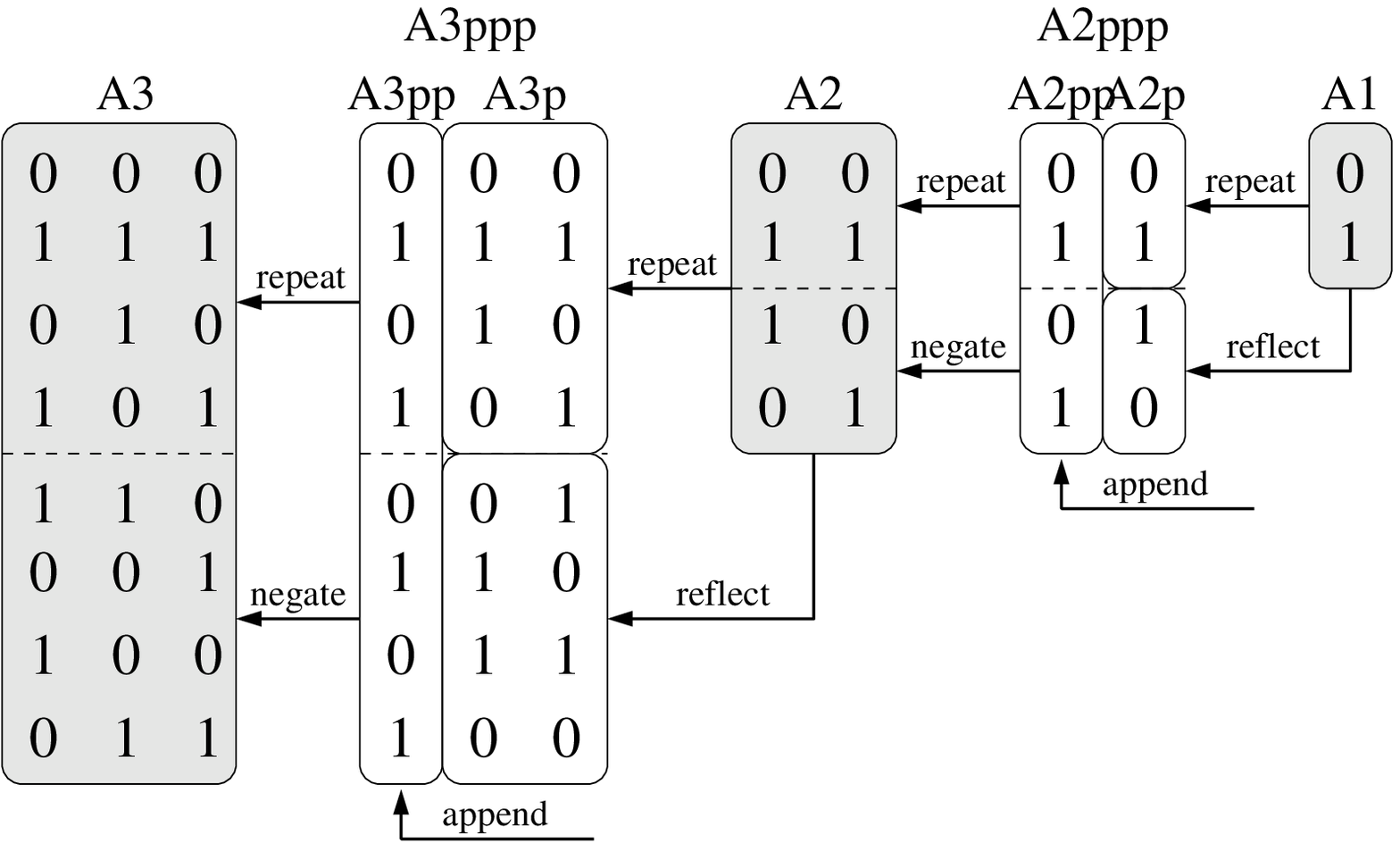}
	\caption[]{Proposed recursive construction of an \gls{agc} for $m=2$ and $m=3$.}
	\label{AGC_construction}
\end{center}
\end{figure*}
This recursive construction is illustrated in \figref{AGC_construction} for $m=2$ and $m=3$. The following lemma shows that it indeed leads to a valid labeling.

\begin{lemma}\label{AGC.Lemma}
All the rows in $\AGC_{m}$ are unique and every odd row differs in $m$ bits compared to its subsequent row.
\end{lemma}
\begin{IEEEproof}
We shall prove \lemmaref{AGC.Lemma} by induction. Let all rows in $\AGC_{m-1}$ be unique and every odd row differ in $m-1$ bits compared to its subsequent row. $\AGC_{1}$ clearly fulfills both criteria. Since the number of rows in $\AGC_{m-1}$ is even, it follows by Step $1$ that every odd row in the upper half of $\AGC'_m$ is identical to an even row in the lower half of $\AGC'_m$, which directly implies that all rows of $\AGC''_m$ in Step $2$ are unique. This also implies that every odd row of $\AGC''_m$ differs in $m$ bits compared to the row below, since the corresponding rows of $\AGC'_m$ differ in $m-1$ bits. Negating all the bits in the lower half of $\AGC''_m$ is therefore equivalent to swapping every odd row in the lower half of $\AGC''_m$ with the row below. Consequently, all rows in $\AGC_m$ are unique and every odd row differs in $m$ bits compared to its subsequent row.
\end{IEEEproof}

We next show that, for $M$PAM constellations, the \gls{agc} maximizes $\Reul$.
\begin{theorem}\label{GC.NBC.AGC.Cel.Theo}
For $\mcX=\mcE$, the \gls{agc} achieves the upper bounds in \eqref{Bounds.Cel} and \eqref{Bounds.Reul}, i.e.,
\begin{align}
\label{CeulAGC}
\CelAGC & = 2mM-2m-M+2\\
\label{ReulAGC}
\ReulAGC & = m-\frac{M-2}{2M-2}.
\end{align}
\end{theorem}
\begin{IEEEproof}
Let $H_m$ denote twice the sum of the Hamming distances between all adjacent rows in $\AGC_m$, and let $H'_m$ and $H''_m$ denote the same quantity for $\AGC'_m$ and $\AGC''_m$, respectively. Recall that every $M$PAM constellation satisfies $H_m=\Cel$. Steps $1$ and $2$ give $H'_m=2H_{m-1}$ and $H''_m=H'_m+2(M-1)$. It then follows that $H_m=H''_m-2+2(m-1)$, since row $M/2$ and row $M/2+1$ in $\AGC''_m$ differ in only one bit and therefore the same rows in $\AGC_m$ differ in $m-1$ bits. This gives $H_m=2H_{m-1}+2(M+m-3)$, which combined with $H_1=2$ gives $H_m=2\left(mM-m-M/2+1\right)$, which proves \eqref{CeulAGC}. Together with \eqref{Rxpl.Final.Uniform} and \eqref{A.Uniform}, this proves \theoref{GC.NBC.AGC.Cel.Theo}.
\end{IEEEproof}

The labeling $\labeling_\tr{AGC}=[0, 3, 2, 1]$ in \exref{Example.4PAM} and \figref{MI_and_MMSE_ratios_4PAM_all_labelings_rhodB} (i.e., $\AGC_2$ in \figref{AGC_construction}) is the \gls{agc} for $4$PAM and gives $\Reul=5/3$.
For $8$PAM, the \gls{agc} is $\labeling_{\tr{AGC}}=[0, 7, 2, 5, 6, 1, 4, 3]$ ($\AGC_3$ in \figref{AGC_construction}) and gives $\Reul=18/7$ as shown in \figref{MMSE_ratios_8PAM_all_labelings_rhodB}.

Revisiting \exref{Example7}, we note that, by \theoref{GC.NBC.AGC.Cel.Theo}, the labeling that achieves the upper bound \eqref{Bounds.Reul} is the \gls{agc} $\AGC_4$. It can be further shown that the labeling with the second largest $\Reul$ can be constructed by reversing the order of the three first rows of the \gls{agc} $\AGC_4$. This demonstrates that for $16$PAM there exist indeed $39$ classes of labelings with different high-\gls{snr} behaviors and hence the bound in \eqref{Size.Cel} is tight for this case.

\section{Conclusions}\label{Sec:Conclusions}

We studied the discrete-time, scalar (real-valued), \gls{awgn} channel when the input takes value in a finite constellation and derived high-\gls{snr} asymptotic expressions for the \gls{mi}, \gls{mmse}, \gls{sep}, the \gls{bicm-gmi}, its derivative, and the \gls{bep}. Our results show that, as the SNR tends to infinity, all these quantities converge to their asymptotes proportionally to $\QF\left({\sqrt{\rho}\MED}/{2}\right)$, where $\MED$ is the \gls{med} of the constellation.

For a uniform input distribution, the proportionality constants for the \gls{mi}, \gls{sep}, and \gls{mmse} were found to be a function of the \gls{med} of the constellation and the number of pairs of constellation points at \gls{med} only. Consequently, the constellation that maximizes the \gls{mi} in the high-SNR regime is the same that minimizes both the \gls{sep} and the \gls{mmse}.

We then applied our results to study binary labelings for \gls{bicm}. By characterizing the high-SNR behavior of the \gls{bicm-gmi}, we proved the long-standing conjecture that Gray codes are optimal at high \gls{snr}. We also proved that there always exists an anti-Gray code for $M$PAM constellations, which is the labeling that has the lowest \gls{bicm-gmi} and the highest \gls{bep} at high \gls{snr}.

\appendices
\section{Proof of \theoref{MI.Gen.Asym.Theo}}\label{MI.Gen.Asym.Theo.Proof}

Both the numerator and the denominator on the \gls{lhs} of \eqref{MI.Gen.Asymptotic} tend to zero as $\rho$ tends to infinity. Thus, it follows from L'H\^{o}pital's rule that 
\begin{align}\label{App.A.1.1}
\limrinf{\frac{\ENX-\MIxp}{\QF\left({\sqrt{\rho}\MED}/{2}\right)}}
& = \limrinf{\frac{\frac{\tr{d}}{\tr{d}\rho}\left(\ENX-\MIxp\right)}{\frac{\tr{d}}{\tr{d}\rho}\QF\left({\sqrt{\rho}\MED}/{2}\right)}}\\
\label{App.A.1.2}
& =
\frac{4}{\MED^2}\limrinf{\frac{\MMSExp}{\GF\left({\sqrt{\rho}\MED}/{2}\right)}}\\
\label{App.A.1.3}
&=
\frac{4}{\MED^2}\limrinf{\frac{\MMSExp}{\QF\left({\sqrt{\rho}\MED}/{2}\right)}}\\
\label{App.A.1.4}
&=
\pi\Bxp
\end{align}
where $\GF(x)$ is defined as
\begin{align}\label{G}
\GF(x) \triangleq \frac{1}{x}\frac{1}{\sqrt{2\pi}}\exp{-\frac{x^2}{2}}.
\end{align}
Here the last step follows from \theoref{MMSE.Gen.Asym.Theo} (proved in \appref{MMSE.Gen.Asym.Theo.Proof}), which also demonstrates that the limit on the \gls{rhs} of \eqref{App.A.1.1} exists. To pass from \eqref{App.A.1.1} to \eqref{App.A.1.2} we used \eqref{dMI_MMSE} and
\begin{align}\label{Q.derivative}
\frac{\tr{d}}{\tr{d}\rho}\QF\left({\sqrt{\rho}\MED}/{2}\right)=-\frac{\MED^2}{8}\GF\left({\sqrt{\rho}\MED}/{2}\right).
\end{align}
To pass from \eqref{App.A.1.2} to \eqref{App.A.1.3} we used \cite[Prop.~19.4.2]{Lapidoth09_Book} to obtain
\begin{align}\label{App.A.2}
\limxinf{\frac{\GF(x)}{\QF(x)}} = 1.
\end{align}
This proves \theoref{MI.Gen.Asym.Theo}.

\section{Proof of \theoref{MMSE.Gen.Asym.Theo}}\label{MMSE.Gen.Asym.Theo.Proof}

For the AWGN channel in \eqref{AWGN}, the conditional \gls{me} is given by \cite[eq.~(22)]{Lozano06}
\begin{align}\label{X.ME}
\hat{X}^{\textME}(y) = \frac{\sumj\Pxj x_j\exp{-\frac{1}{2}(\by-\sqrt{\rho}x_j)^2}}{\sum_{j\in\mcIX}\Pxj\exp{-\frac{1}{2}(\by-\sqrt{\rho}x_j)^2}}.
\end{align}
By using \eqref{X.ME} in \eqref{mmse-def}, we obtain
\begin{align}\label{MMSE.Proof.1}
\MMSExp & =
\sumi
\Pxi
\int_{-\infty}^{\infty}
\frac{1}{\sqrt{2\pi}}
\exp{-\frac{1}{2}(\by-\sqrt{\rho}x_i)^2} \nonumber\\
&
\cd\left(
\frac{\sum_{j\in\mcIX}\Pxj(x_i-x_j)\exp{-\frac{1}{2}(\by-\sqrt{\rho}x_j)^2}}{\sum_{j\in\mcIX}\Pxj\exp{-\frac{1}{2}(\by-\sqrt{\rho}x_j)^2}}
\right)^2
\,\tr{d}\by\\
\label{MMSE.Proof.2}
& =
\sumi
\Pxi
V_i(\rho)
\end{align}
where
\begin{align}
V_i(\rho) & \triangleq
\int_{-\infty}^{\infty}
\frac{\exp{-t^2}}{\sqrt{\pi}}
\left(
\frac{
\sum_{\delta\in\mcDxi} \delta \Rxi(\delta)\cd\exp{-\sqrt{2\rho}{t}{\delta}-\frac{\rho\delta^2}{2}}
}{
\sum_{\delta\in\mcDxi} \Rxi(\delta)\cd\exp{-\sqrt{2\rho}{t}{\delta}-\frac{\rho\delta^2}{2}}
}
\right)^2
\,\tr{d}t
\label{Ii.Def.MMSE}
\end{align}
with
$\mcDxi \triangleq\set{\delta : \delta=x_i-x, x\in\mcX}$
and
\begin{align}
\label{R.xi.def}
\Rxi(\delta)
&\triangleq
\begin{cases}
\frac{\Pxj}{\Pxi}, 		& \text{if $\exists x_j\in\mcX: x_i-x_j=\delta$}\\
0, 					& \text{otherwise}
\end{cases}
\end{align}
and where to pass from \eqref{MMSE.Proof.1} to \eqref{MMSE.Proof.2} we used the substitution $\by-\sqrt{\rho}x_i=\sqrt{2}t$.

Combining \eqref{App.A.2} and \eqref{MMSE.Proof.2}, we obtain
\begin{align}
\limrinf{
\frac{\MMSExp}{\QF\left({\sqrt{\rho}\MED}/{2}\right)}
}
\label{App.B.1.5}
&=
\sumi\Pxi \limrinf{\frac{V_i(\rho)}{\GF\left({\sqrt{\rho}\MED}/{2}\right)}}.
\end{align}
As will become apparent later, the limit on the \gls{rhs} of \eqref{App.B.1.5} exists and, hence, so does the limit on the \gls{lhs}

Using \eqref{Ii.Def.MMSE} and \eqref{G}, and the substitution $r=d\sqrt{\rho/8}$, we obtain
\begin{align}\label{App.B.3}
\limrinf{
\frac{V_i(\rho)}
{
\GF\left({\sqrt{\rho}d}/{2}\right) }
}
=
2
\left(\limrrinf{F_{i}^{-}(r)}+\limrrinf{F_{i}^{+}(r)}\right)
\end{align}
where
\begin{align}
\nonumber
&F_{i}^{-}(r) \triangleq\\
\label{App.B.Fm}
&
\int_{-\infty}^{0}
r\exp{r^2-t^2}
\left(
\frac{\sum_{\delta\in\mcDxi}\delta \Rxi(\delta)\exp{-4rt\frac{\delta}{d}-4r^2\frac{\delta^2}{d^2}}}{\sum_{\delta\in\mcDxi}\Rxi(\delta)\exp{-4rt\frac{\delta}{d}-4r^2\frac{\delta^2}{d^2}}}
\right)^2
\,\tr{d}t
\end{align}
and
\begin{align}
\nonumber
&F_{i}^{+}(r) \triangleq \\
\label{App.B.Fp}
&
\int_{0}^{\infty}
r\exp{r^2-t^2}
\left(
\frac{\sum_{\delta\in\mcDxi}\delta \Rxi(\delta)\exp{-4rt\frac{\delta}{d}-4r^2\frac{\delta^2}{d^2}}}{\sum_{\delta\in\mcDxi}\Rxi(\delta)\exp{-4rt\frac{\delta}{d}-4r^2\frac{\delta^2}{d^2}}}
\right)^2
\,\tr{d}t.
\end{align}

We will next calculate the first limit in \eqref{App.B.3}. Using the substitution $t=u/r-r$ we express $F_{i}^{-}(r)$ as
\begin{align}
{F_{i}^{-}(r)}
\label{App.B.Fm.2}
&=
{
\int_{-\infty}^{\infty}
f_{i}^{-}(r,u)
\,\tr{d}u}
\end{align}
where
\begin{align}
\nonumber
f_{i}^{-}&(r,u) \triangleq h(r^2-u)\cd \exp{2u-\frac{u^2}{r^2}}\\
&
\qquad \cd \left(
\frac{\sum_{\delta\in\mcD_i^*} \delta \Rxi(\delta)\exp{-4u\frac{\delta}{d}-4r^2U(\delta)}}{1+\sum_{\delta\in\mcD_i^*}\Rxi(\delta)\exp{-4u\frac{\delta}{d}-4r^2U(\delta)}}
\right)^2
\label{App.B.fim}
\end{align}
with $\mcD_i^*\triangleq \mcDxi\setminus\set{0}$,
\begin{align}\label{App.A.K2}
U(\delta) \triangleq \frac{\delta}{d}\left(\frac{\delta}{d}-1\right)
\end{align}
and $h(x)$ being Heaviside's step function (i.e., $h(x)=1$ if $x\geq 0$ and $h(x)=0$ if $x<0$). Using the fact that $U(\delta)\geq 0, \forall \delta\in\mcD_i^*$ and $U(\MED)=0$, we obtain
\begin{align}
\label{App.B.Fm.2.5}
\limrrinf{
f_{i}^{-}(r,u)
}
& =
\MED^{2}
\exp{2u}
{
\left(
\frac{
 \Rxi(\MED)\exp{-4u}
}
{1+\Rxi(\MED)\exp{-4u}}
\right)^2
}.
\end{align}
As we shall prove in \lemmaref{App.B.fim.DCT} ahead, $u\mapsto f_i^{-}(r,u)$ is uniformly bounded by some integrable function $u \mapsto g_i^{-}(u)$ that is independent of $r$. It thus follows from Lebesgue's Dominated Convergence Theorem \cite[Theorem~1.34]{Rudin87_Book} that
\begin{align}
\label{App.B.Fm.3}
\limrrinf{F_{i}^{-}(r)} & =
\int_{-\infty}^{\infty}
\limrrinf{
f_{i}^{-}(r,u)}
\,\tr{d}u\\
\label{App.B.Fm.4}
&  = \frac{\MED^2\sqrt{\Rxi(\MED)}}{2}
\int_{0}^{\infty}
{
\frac{\xi^{2}}{(1+\xi^{2})^{2}}
}
\,\tr{d}\xi\\
\label{App.B.Fm.final}
 & = \frac{\MED^2 \pi \sqrt{\Rxi(\MED)}}{8}
\end{align}
where \eqref{App.B.Fm.4} follows from \eqref{App.B.Fm.2.5} and the substitution $\sqrt{\Rxi(\MED)}\exp{-2u}=\xi$, and \eqref{App.B.Fm.final} follows from \cite[eq.~(3.241.5)]{Gradshteyn00_Book}.

It thus remains to show that $u\mapsto f_i^{-}(r,u)$ is uniformly bounded by some integrable function $u \mapsto g_i^{-}(u)$ that is independent of $r$. We do this in the following lemma.

\begin{lemma}\label{App.B.fim.DCT}
For any $r>0$
\begin{equation}\label{App.B.fim.DCT.Lemma.eq}
0 \leq f_i^{-}(r,u)\leq g_i^{-}(u), \quad u\in\Real
\end{equation}
where
\begin{align}\label{App.B.gim}
g_{i}^{-}(u) \triangleq \frac{\MaxED^2(M-1)^2}{\Pxi^2}\exp{-2|u|}
\end{align}
and $\MaxED$ is the maximum Euclidean distance of the constellation, \ie $\MaxED \triangleq \max_{i,j\in\mcIX}|x_i-x_j|$. Furthermore,
\begin{align}
\int_{-\infty}^{\infty} g_{i}^{-}(u)\,\tr{d}u = \frac{\MaxED^2(M-1)^2}{\Pxi^2} < \infty.
\end{align}
\end{lemma}
\begin{IEEEproof}
We first note that $f_i^{-}(r,u)\geq 0$, $r>0$, $u\in\Real$. It thus remains to show the second inequality in \eqref{App.B.fim.DCT.Lemma.eq}. To this end, we use $h(r^2-u)\le 1$, $\exp{-\frac{u^2}{r^2}}\le 1$, and $\delta \le \MaxED $ to upper-bound \eqref{App.B.fim} as
\begin{align}
f_{i}^{-}(r,u)
& \leq \MaxED^2
\exp{2u}
\left(
\frac{\sum_{\delta\in\mcD_i^*} \Rxi(\delta)\exp{-4u\frac{\delta}{\MED}-4r^2U(\delta)}}{1+\sum_{\delta\in\mcD_i^*}\Rxi(\delta)\exp{-4u\frac{\delta}{\MED}-4r^2U(\delta)}}
\right)^2 \\
\label{App.B.fim.4}
& = \MaxED^2 \exp{2u}
\left(1+\frac{1}{\sum_{\delta\in\mcD_i^*}\Rxi(\delta)\exp{-4u\frac{\delta}{\MED}-4r^2U(\delta)}}
\right)^{-2}.
\end{align}
Since $\Rxi(\delta)<1/\Pxi$ and $\exp{-4r^2U(\delta)}\le 1$, we can further upper-bound \eqref{App.B.fim.4} as
\begin{align}
\label{App.B.fim.4.5}
f_{i}^{-}(r,u)
& <
\MaxED^2 \exp{2u}
\left(1+\frac{\Pxi}{\sum_{\delta\in\mcD_i^*}\exp{-4u\frac{\delta}{\MED}}}
\right)^{-2}\\
\label{App.B.fim.5}
& <  \frac{\MaxED^2 \exp{2u}}{\Pxi^2}
\left(1+\frac{1}{\sum_{\delta\in\mcD_i^*}\exp{-4u\frac{\delta}{\MED}}}
\right)^{-2}
\end{align}
where to pass from \eqref{App.B.fim.4.5} to \eqref{App.B.fim.5} we used $\Pxi<1$.

For $u\geq 0$, we have
\begin{align}
\label{App.B.fim.8}
f_{i}^{-}(r,u) & <
\frac{\MaxED^2 \exp{2u}}{\Pxi^2}
\left(1+\frac{1}{(M-1)\exp{-4u}}
\right)^{-2}\\
\label{App.B.fim.9}
& < \frac{\MaxED^2 (M-1)^2}{\Pxi^2}\exp{-6u}\\
\label{App.B.fim.11}
& < \frac{\MaxED^2 (M-1)^2}{\Pxi^2}\exp{-2|u|}
\end{align}
where to pass from \eqref{App.B.fim.5} to \eqref{App.B.fim.8} we upper-bounded the $(M-1)$ exponentials in the summation by $\exp{-4u}$.

For $u\leq 0$, we have
\begin{align}
\label{App.B.fim.12}
f_{i}^{-}(r,u) & < \frac{\MaxED^2 \exp{2u}}{\Pxi^2} \\
		& \le \frac{\MaxED^2 (M-1)^2 }{\Pxi^2} \exp{-2|u|} \label{App.B.fim.13}		
\end{align}
where \eqref{App.B.fim.12} follows from discarding the sum of exponentials in \eqref{App.B.fim.5}. Combining \eqref{App.B.fim.11} and \eqref{App.B.fim.13} gives \eqref{App.B.gim}. This proves \lemmaref{App.B.fim.DCT}.
\end{IEEEproof}

Returning to the proof of \theoref{MMSE.Gen.Asym.Theo}, the second limit on the \gls{rhs} of \eqref{App.B.3} can be computed along the same lines by using the substitution $t=u/r+r$ in \eqref{App.B.Fp}:
\begin{align}
\label{App.B.Fp.final.m1}
\limrrinf{F_{i}^{+}(r)} & = \frac{\MED^2 \pi \sqrt{\Rxi(-\MED)}}{8}.
\end{align}
Using \eqref{App.B.Fm.final} and \eqref{App.B.Fp.final.m1} in \eqref{App.B.3}, and combining the result with \eqref{App.B.1.5} yields
\begin{equation}
\limrinf{
\frac{\MMSExp}{\QF\left({\sqrt{\rho}\MED}/{2}\right)}
} = \sumi\Pxi \frac{\pi d^2}{4}\left(\sqrt{\Rxi(\MED)}+\sqrt{\Rxi(-\MED)}\right)
\end{equation}
which in view of \eqref{R.xi.def} and \eqref{B} is equal to $\pi d^2\Bxp/4$. This proves \theoref{MMSE.Gen.Asym.Theo}.


\section{Proof of \theoref{SEP.Gen.Asym.Theo}}\label{SEP.Gen.Asym.Theo.Proof}

Using Bayes' rule, $\hat{X}^{\textMAP}(y)$ in \eqref{map} can be expressed as
\begin{align}\label{SEP.Proof.map.1}
\hat{X}^{\textMAP}(y)&= \argmax_{x\in\mcX} \left\{\pdf_{Y|X}(y|x)\pmf_X(x)\right\}\\
\label{SEP.Proof.map.2}
& =x_j, \quad \text{if } y\in \mcY_j(\rho)
\end{align}
where $\mcY_j(\rho)$ is the decision region for the symbol $x_j$ with $j=1,\ld,M$. For sufficiently large $\rho$, these decision regions can be written as
\begin{align}
\mcY_j(\rho)
\label{SEP.Proof.Voronoi.2}
& \triangleq \set{y\in\Real: \beta_{j-1}(\rho) \le y < \beta_j(\rho)}
\end{align}
where $\beta_{\ell}(\rho)$ with $\ell=0,\dots,M$ are the $M+1$ thresholds defining the $M$ regions, \ie
\begin{align}\label{SEP.Proof.beta.MAP}
\beta_{\ell}(\rho)=
\begin{cases}
-\infty, 	& \text{$\ell=0$}\\
\frac{\log(p_{\ell}/p_{\ell+1})}{\sqrt{\rho}(x_{\ell+1}-x_{\ell})}+\frac{\sqrt{\rho}(x_{\ell+1}+x_{\ell})}{2}, 	 & \text{$\ell=1,\ld,M-1$}\\
+\infty, 				& \text{$\ell=M$}\\
\end{cases}
\end{align}
which are obtained by solving
\begin{align}\label{SEP.Proof.threshold}
p_{\ell} \pdf_{Y|X}(\beta_{\ell}(\rho)|x_{\ell})=p_{\ell+1} \pdf_{Y|X}(\beta_{\ell}(\rho)|x_{\ell+1}).
\end{align}

The following lemma will be used in this proof as well as in the proof of \theoref{BEP.Gen.Asym.Theo} (\appref{BEP.Gen.Asym.Theo.Proof}).
\begin{lemma}\label{beta.Lemma}
For any $\PMF$ and $i\in\mcIX$
\begin{align}
\nonumber
\limrinf{
\frac{\QF\left(|\beta_{\ell}(\rho)-\sqrt{\rho}x_i|\right)}{\QF\left(\sqrt{\rho}\MED/2\right)}
} \\
\label{beta.Lemma.eq}
& \hspace{-2cm} =
\begin{cases}
\sqrt{\Rxi(\MED)}, 	& \text{if $\ell=i-1$}\\
\sqrt{\Rxi(-\MED)}, 	& \text{if $\ell=i$}\\
0, 				& \text{if $\ell\notin\set{i-1,i}$}\\
\end{cases}
\end{align}
where $\beta_{\ell}(\rho)$ is given by \eqref{SEP.Proof.beta.MAP} and $\Rxi(\delta)$ by \eqref{R.xi.def}.
\end{lemma}
\begin{IEEEproof}
We use \eqref{SEP.Proof.beta.MAP} to obtain
\begin{align}
\label{beta.Lemma.Proof.1}
\beta_{\ell}(\rho)-\sqrt{\rho}x_i & =
\frac{\log(p_{\ell}/p_{\ell+1})}{\sqrt{\rho}(x_{\ell+1}-x_{\ell})}+\frac{\sqrt{\rho}\tdelta_{i,\ell}}{2}
\end{align}
where for any $i,\ell$
\begin{align}
\label{beta.Lemma.Proof.tdelta}
\tdelta_{i,\ell} & \triangleq x_{\ell+1}+x_{\ell}-2x_{i}.
\end{align}
We form the ratio
\begin{align}
\nonumber
&\frac{\GF\left(|\beta_{\ell}(\rho)-\sqrt{\rho}x_i|\right)}{\GF\left(\sqrt{\rho}\MED/2\right)}
 = \frac{\rho \MED(x_{\ell+1}-x_{\ell})}{|2\log({p_{\ell}}/{p_{\ell+1}})+\rho\tdelta_{i,\ell}(x_{\ell+1}-x_{\ell})|}\\
\label{beta.Lemma.Proof.2}
&\cd
\mathrm{exp}\left(
-\frac{\bigl(\log{\frac{p_{\ell}}{p_{\ell+1}}}\bigr)^2}{2\rho(x_{\ell+1}-x_{\ell})^2}
-\frac{\tdelta_{i,\ell}\log{\frac{p_{\ell}}{p_{\ell+1}}}}{2(x_{\ell+1}-x_{\ell})}
-\frac{\rho(\tdelta_{i,\ell}^2-\MED^2)}{8}
\right)
\end{align}
and study the limit
\begin{align}\label{limit.Lemma.Proof}
\limrinf{\frac{\GF\left(|\beta_{\ell}(\rho)-\sqrt{\rho}x_i|\right)}{\GF\left(\sqrt{\rho}\MED/2\right)}}.
\end{align}
To this end, we distinguish between three cases:
\begin{enumerate}[(i)]
\item If $i=\ell$ and $x_{\ell+1}-x_{\ell}=d$, then $\tdelta_{i,\ell}=x_{\ell+1}-x_{\ell}=\MED$ and the limit in \eqref{limit.Lemma.Proof} is $\exp{-\log{(p_{\ell}/p_{\ell+1})/2}}=\sqrt{p_{\ell+1}/p_{\ell}}$.
\item If $i=\ell+1$ and $x_{\ell+1}-x_{\ell}=\MED$, then $\tdelta_{i,\ell}=x_{\ell}-x_{\ell+1}=-\MED$ and the limit in \eqref{limit.Lemma.Proof} is $\sqrt{p_{\ell}/p_{\ell+1}}$.
\item In all other cases, $|\tdelta_{i,\ell}|>\MED$ and the limit in \eqref{limit.Lemma.Proof} is zero.
\end{enumerate}

A slight change in notation then yields
\begin{align}
\nonumber
&\limrinf{\frac{\GF\left(|\beta_{\ell}(\rho)-\sqrt{\rho}x_i|\right)}{\GF\left(\sqrt{\rho}\MED/2\right)}}\\
\label{beta.Lemma.Proof.end}
&\qquad =
\begin{cases}
\sqrt{\frac{p_{i+1}}{p_{i}}}, 	& \text{if $\ell=i$ and $x_{\ell+1}-x_{\ell}=\MED$}\\
\sqrt{\frac{p_{i-1}}{p_{i}}}, 	& \text{if $\ell=i-1$ and $x_{\ell+1}-x_{\ell}=\MED$}\\
0, 					& \text{otherwise}\\
\end{cases}.
\end{align}
Combining \eqref{beta.Lemma.Proof.end} with \eqref{R.xi.def} and \eqref{App.A.2} proves \lemmaref{beta.Lemma}.
\end{IEEEproof}

Returning to the proof of \theoref{SEP.Gen.Asym.Theo}, using \eqref{SEP.Proof.map.2} and \eqref{SEP.Proof.Voronoi.2}, the \gls{sep} in \eqref{sep} can be written as
\begin{align}
\label{SEP.Proof.SEP.1}
\SEPxp 			& = \sumi \Pxi \Pr\set{Y\notin \mcY_i(\rho)| X=x_i}\\
\label{SEP.Proof.SEP.2}
				& = \sumi \Pxi\Bigl(\QF\left(\beta_i(\rho)-\sqrt{\rho}x_i\right) \nonumber\\
				&\qquad\qquad\quad+ \QF\left(\sqrt{\rho}x_{i}-\beta_{i-1}(\rho)\right)\Bigr)
\end{align}
which gives
\begin{align}
\notag
\label{SEP.Proof.SEP.3}
\limrinf{
\frac{\SEPxp}{\QF({\sqrt{\rho} \MED}/{2})}
}
& =
\sumi \Pxi
\biggl(
\limrinf{
\frac{\QF\left(\beta_i(\rho)-\sqrt{\rho}x_i\right)}{\QF({\sqrt{\rho} \MED}/{2})}
}
\\
&\qquad\,+
\limrinf{
\frac{\QF\left(\sqrt{\rho}x_{i}-\beta_{i-1}(\rho)\right)}{\QF({\sqrt{\rho} \MED}/{2})}
}
\biggr)\\
\label{SEP.Proof.SEP.4}
&=
\sumi \Pxi
\biggl(
\sqrt{\Rxi(-\MED)}+\sqrt{\Rxi(\MED)}
\biggr)
\end{align}
where to pass from \eqref{SEP.Proof.SEP.3} to \eqref{SEP.Proof.SEP.4} we used \lemmaref{beta.Lemma} twice, observing that the arguments of both Q-functions are positive for large enough $\rho$. The proof of \theoref{SEP.Gen.Asym.Theo} is completed by combining \eqref{SEP.Proof.SEP.4} with \eqref{B} and \eqref{R.xi.def}.

\section{Proof of \theoref{BICM-GMI.Asym.Theo}}\label{BICM-GMI.Asym.Proof}

Using the expression for the \gls{bicm-gmi} \eqref{BICM.GMI.General}, we have
\begin{align}
\nonumber
\ENX & -\MIBIxp\\
\nonumber
 & = \quad
\sumk
(\ENX-\MIxp)\\
&\quad-\sumk\sumb P_{Q_k}(b)(\ENXxpkb-\MIxpkb)\nonumber\\
 & \quad-(m-1) \ENX+\sumk\sumb P_{Q_k}(b) \ENXxpkb.
\label{BICM-GMI.Asym.proof.1}
\end{align}
The last two terms on the \gls{rhs} of \eqref{BICM-GMI.Asym.proof.1} cancel because
\begin{align}
\sumk & \sumb \pmf_{Q_k}(b)\ENXxpkb\nonumber\\
& = -\sumk\sumb\sumikb \pmf_{Q_k}(b)\PMFkb(x_i)\log{\PMFkb(x_i)}\label{BICM-GMI.Asym.proof.2}\\
& = -\sumk\sumi \Pxi \log \frac{\Pxi}{\pmf_{Q_k}(q_{i,k})}\label{BICM-GMI.Asym.proof.3}\\
& = m \ENX +\sumi\Pxi\sumk\log{\pmf_{Q_k}(q_{i,k})}\label{BICM-GMI.Asym.proof.4}\\
& = m \ENX + \sumi\Pxi\log{\prod_{k=1}^{m}\pmf_{Q_k}(q_{i,k})}\label{BICM-GMI.Asym.proof.5}\\
& = m \ENX - \ENX \label{BICM-GMI.Asym.proof.6}
\end{align}
where to pass from \eqref{BICM-GMI.Asym.proof.2} to \eqref{BICM-GMI.Asym.proof.3} we used \eqref{Pxi.Cku}, and to pass from \eqref{BICM-GMI.Asym.proof.5} to \eqref{BICM-GMI.Asym.proof.6} we used \eqref{Pxi.BICM}.

We divide both sides of \eqref{BICM-GMI.Asym.proof.1} by $\QF(\sqrt{\rho}\MED/2)$ and take the limit as $\rho \rightarrow \infty$. For the first two terms, we change the order of summation and limit and apply \theoref{MI.Gen.Asym.Theo} to each term. This gives
\begin{align}
\nonumber
&\limrinf{
\frac{\ENX-\MIBIxp}{\QF\left({\sqrt{\rho}\MED}/{2}\right)}
}  \\
\label{BICM-GMI.Asym.proof.7}
&\qquad = \pi \sumk \Biggl(\Bxp - \sumb \pmf_{Q_k}(b)\Bxpkb\Biggr).
\end{align}
\theoref{BICM-GMI.Asym.Theo} follows by showing that the \gls{rhs} of \eqref{BICM-GMI.Asym.proof.7} is equal to $\pi \Dxl$. We shall do this in the following lemma, which will also be used in the proof of \theoref{BICM-GMMSE.Asym.Theo}.

\begin{lemma}\label{lemma.sumk.Dxl}
We have
\begin{align}\label{lemma.sumk.Dxl.eq}
\sumk \Biggl(\Bxp - \sumb \pmf_{Q_k}(b)\Bxpkb\Biggr)
= \Dxl
\end{align}
where $\Bxp$ is given by \eqref{B}, $\Dxl$ by \eqref{D},
\begin{align}\label{B.subconstellation}
\Bxpkb \triangleq \sumikb\sum_{\substack{j\in\mcIXkb\\|x_{i}-x_{j}|=\MED}} \sqrt{\PMFkb(x_i)\PMFkb(x_j)}
\end{align}
and $\PMFkb(x)$ is given by \eqref{Pxi.Cku}.
\end{lemma}
\begin{IEEEproof}
Express the inner sum on the \gls{lhs} of \eqref{lemma.sumk.Dxl.eq} using \eqref{Pxi.Cku} and \eqref{B.subconstellation} as
\begin{align}\label{Lemma7.proof.1}
\sumb \pmf_{Q_k}(b)\Bxpkb = \sumb \sumikb\sum_{\substack{j\in\mcIXkb\\|x_{i}-x_{j}|=\MED}} \sqrt{\Pxj\Pxi}.
\end{align}
Expanding the sum in \eqref{B} using $\mcX=\mcXkz\cup\mcXko$, we obtain
\begin{align}
\nonumber
\Bxp &= \sumb \sumikb\sum_{\substack{j\in\mcIXkb\\|x_{i}-x_{j}|=\MED}} \sqrt{\Pxj\Pxi} \\
\label{Bxp.expanded}
&\qquad + 2\sum_{i\in\mcIXkz}\sum_{\substack{j\in\mcIXko\\|x_{i}-x_{j}|=\MED}} \sqrt{\Pxj\Pxi}.
\end{align}
\lemmaref{lemma.sumk.Dxl} follows by applying \eqref{Lemma7.proof.1} and \eqref{Bxp.expanded} to the \gls{lhs} of \eqref{lemma.sumk.Dxl.eq} and by using the definition of $\Dxl$ in \eqref{D}.
\end{IEEEproof}

\section{Proof of \theoref{BICM-GMMSE.Asym.Theo}}\label{BICM-GMMSE.Asym.Theo.Proof}

We divide the \gls{lhs} of \eqref{BICM.GMMSE.General.0} and \eqref{BICM.GMMSE.General} by $\QF(\sqrt{\rho}\MED/2)$ and take the limit as $\rho\to\infty$ to obtain
\begin{align}
\nonumber
&\limrinf{
\frac{\MMSEBIxp}{\QF\left({\sqrt{\rho}\MED}/{2}\right)}
} \\
&
\,
=
\limrinf{
\sumk\Biggl(\frac{\MMSExp}{\QF\left({\sqrt{\rho}\MED}/{2}\right)}-\sumb \pmf_{Q_k}(b)\frac{\MMSEkpkb}{\QF\left({\sqrt{\rho}\MED}/{2}\right)}\Biggr)
}.
\end{align}

Changing the order of summation and limit, and applying \theoref{MMSE.Gen.Asym.Theo} yields
\begin{align}
\limrinf{
\frac{\MMSEBIxp}{\QF\left({\sqrt{\rho}\MED}/{2}\right)}
}
\label{App.BICM-GMMSE.1}
= \frac{\pi\MED^2}{4} \sumk \Biggl(\Bxp - \sumb \pmf_{Q_k}(b)\Bxpkb\Biggr)
\end{align}
where $\Bxpkb$ is given by \eqref{B.subconstellation}. \theoref{BICM-GMMSE.Asym.Theo} follows by noting that, by \lemmaref{lemma.sumk.Dxl}, the \gls{rhs} of \eqref{App.BICM-GMMSE.1} is equal to $\frac{\pi\MED^{2}}{4} \Dxl$.

\section{Proof of \theoref{BEP.Gen.Asym.Theo}}\label{BEP.Gen.Asym.Theo.Proof}

Using the law of total probability, the \gls{bep} in \eqref{bep} can be written as
\begin{align}
\notag
& \BEPxp 	 \\
\label{BEP.Proof.1}
& = \frac{1}{m}\sumk \sumb \sumikb \Pxi \Pr\set{\hat{Q}_k^{\textMAP}(Y)\ne q_{k,i}|X=x_i}\\
\label{BEP.Proof.2}
		& = \frac{1}{m}\sumk \sumb \sumikb  \Pxi \Pr\Biggl\{Y\in\bigcup_{j\in\mcIXkbn} \mcY_{j}(\rho)\Bigg|X=x_i\Biggr\}
\end{align}
with $\mcY_j(\rho)$ given by \eqref{SEP.Proof.Voronoi.2} and were we use $\overline{b}$ to denote the negation of a bit $b$. Using the fact that $\mcY_{j}(\rho)$ are disjoint, we rewrite \eqref{BEP.Proof.2} as
\begin{align}
\notag
& \BEPxp 	 \\
\label{BEP.Proof.3}
		& = \frac{1}{m}\sumk \sumb \sumikb  \Pxi \sum_{j\in\mcIXkbn} \Pr\bigl\{Y\in \mcY_{j}(\rho)|X=x_i\bigr\}\\
\label{BEP.Proof.4}
		& = \frac{1}{m}\sumk \sumb \sumikb  \Pxi \sum_{j\in\mcIXkbn} \Gamma_{i,j}(\rho) \\
\nonumber
		& = \frac{1}{m}\sumk \sumb \sumikb  \Pxi \Biggl(\sum_{\substack{j\in\mcIXkbn:j<i}} \Gamma_{i,j}(\rho)\\
\label{BEP.Proof.5}
		& \qquad\,\qquad\qquad\qquad\qquad\qquad {} +\sum_{\substack{j\in\mcIXkbn:j>i}} \Gamma_{i,j}(\rho)\Biggr)
\end{align}
where
\begin{align}\label{BEP.PEP}
\Gamma_{i,j}(\rho)
& \triangleq \QF\left(\beta_{j-1}(\rho)-\sqrt{\rho}x_i\right)-\QF\left(\beta_{j}(\rho)-\sqrt{\rho}x_i\right) \\
& = \QF\left(\sqrt{\rho}x_i-\beta_{j}(\rho)\right)-\QF\left(\sqrt{\rho}x_i-\beta_{j-1}(\rho)\right)\label{BEP.PEP.neg}
\end{align}
and where we have used that $\QF(-x)=1-\QF(x)$.

By using \eqref{BEP.PEP} and \eqref{BEP.PEP.neg} in \eqref{BEP.Proof.5}, dividing both sides of \eqref{BEP.Proof.5} by $\QF\left(\sqrt{\rho}\MED/2\right)$, and taking the limit as $\rho\rightarrow\infty$, we obtain
\begin{align}
\limrinf{
\frac{\BEPxp}{\QF\left(\sqrt{\rho}\MED/2\right)}
}
\label{BEP.Proof.5.5} & =\frac{1}{m}\sumk \sumb \sumikb  \Pxi \sum_{\ell=1}^{4}s_{\ell}
\end{align}
where
\begin{align}
\label{BEP.Proof.5.S1}s_1 &= \sum_{j\in\mcIXkbn: j<i}
\limrinf{\frac{\QF\left(\sqrt{\rho}x_i-\beta_{j}(\rho)\right)}{\QF\left(\sqrt{\rho}\MED/2\right)}}\\
\label{BEP.Proof.5.S2}s_2 &= - \sum_{j\in\mcIXkbn: j<i}
\limrinf{\frac{\QF\left(\sqrt{\rho}x_i-\beta_{j-1}(\rho)\right)}{\QF\left(\sqrt{\rho}\MED/2\right)}}\\
\label{BEP.Proof.5.S3} s_3 &=\sum_{j\in\mcIXkbn: j>i}
\limrinf{\frac{\QF\left(\beta_{j-1}(\rho)-\sqrt{\rho}x_i\right)}{\QF\left(\sqrt{\rho}\MED/2\right)}}\\
\label{BEP.Proof.5.S4}
s_4 &=-\sum_{j\in\mcIXkbn: j>i}
\limrinf{\frac{\QF\left(\beta_{j}(\rho)-\sqrt{\rho}x_i\right)}{\QF\left(\sqrt{\rho}\MED/2\right)}}.
\end{align}
Note that, for sufficiently large $\rho$, the arguments of the Q-functions in \eqref{BEP.Proof.5.S1}--\eqref{BEP.Proof.5.S4} are positive.

By \lemmaref{beta.Lemma}, $s_2=s_4=0$. Furthermore, applying \lemmaref{beta.Lemma} to \eqref{BEP.Proof.5.S1} and \eqref{BEP.Proof.5.S3}, we conclude that the only nonzero contribution to $s_1$ and $s_3$ can come from the terms $j=i-1$ and $j=i+1$, respectively. We therefore express $s_1$ as
\begin{align}\label{S1.nz}
s_1 & =
\begin{cases}
\sqrt{\frac{\Pxim}{\Pxi}}, 		& \text{if $\exists x_{i-1}\in\mcXkbn: x_i-x_{i-1}=\MED$}\\
0, 					& \text{otherwise}
\end{cases}
\end{align}
and $s_3$ as
\begin{align}\label{S3.nz}
s_3 & =
\begin{cases}
\sqrt{\frac{\Pxip}{\Pxi}}, 	& \text{if $\exists x_{i+1}\in\mcXkbn: x_{i+1}-x_{i}=\MED$}\\
0, 					& \text{otherwise}
\end{cases}.
\end{align}
Using \eqref{S1.nz} and \eqref{S3.nz} in \eqref{BEP.Proof.5.5}, we obtain
\begin{align}
\nonumber
\limrinf{
\frac{\BEPxp}{\QF\left(\sqrt{\rho}\MED/2\right)}
}
&\\
\label{BEP.Proof.7}
&
\hspace{-1cm}
=
\frac{1}{m}\sumk \sumb \sumikb   \Pxi \sum_{\substack{j\in\mcIXkbn\\|x_{i}-x_{j}|=d}}  \sqrt{\frac{\Pxj}{\Pxi}}.
\end{align}
The proof is completed by moving $\Pxi$ to the inner sum in \eqref{BEP.Proof.7} and by comparing the resulting expression with \eqref{D}.

\section*{Acknowledgment}\label{Sec:Ack}

The authors wish to thank the Associate Editor Robert F. H. Fischer and the anonymous reviewers for their valuable comments.

\balance

\bibliography{IEEEabrv,references_all}
\bibliographystyle{IEEEtran}

\end{document}